\documentclass[12pt]{article}
\usepackage{epsfig, epsf, graphicx, subfigure}
\usepackage{pstricks, pst-node, psfrag}
\usepackage{amssymb,amsmath,bm}
\usepackage{verbatim,enumerate}
\usepackage{rotating, lscape}
\usepackage{setspace}
\usepackage[english]{babel}
\usepackage[sort&compress]{natbib}
\usepackage{multirow}
\usepackage{theorem}

\def\EV{\mathrm{EV}}
\def\df{\mathrm{d}}

\def\p{\partial}

\def\00{\mathrm{0}}
\def\UU{\mathbf{U}}

\def\ZZ{\mathbf{Z}}

\def\WW{\boldsymbol{W}}
\def\ww{\boldsymbol{w}}
\def\tht{\boldsymbol{\theta}}

\def\uu{\mathbf{u}}

\def\Sig{\boldsymbol{\Sigma}}

\newtheorem{prop}{Proposition}

\begin{document}

\thispagestyle{empty} \baselineskip=28pt \vskip 5mm
\begin{center} {\Huge{\bf Conditional Normal Extreme-Value Copulas }}
\end{center}

\baselineskip=12pt \vskip 10mm

\begin{center}\large
%Ganggang Xu\footnote[1]{
%\baselineskip=10pt Department of Mathematical Sciences,
%Binghamton University, Binghamton, NY 13902, USA. \\
%E-mail: gang@stat.tamu.edu}
Pavel Krupskii\footnote[1]{\baselineskip=10pt University of Melbourne, Parkville 3010, Australia. E-mail: pavel.krupskiy@unimelb.edu.au \\
This research was supported by the Andrew Sisson fund.} 
%Harry Joe\footnote[2]{\baselineskip=10pt University of British Columbia, Vancouver V6T 1Z4, Canada. E-mail: harry.joe@ubc.ca. \\
%This research was supported by NSERC.}
and Marc G. Genton\footnote[2]{\baselineskip=10pt Statistics Program, King Abdullah University of Science and Technology (KAUST),
Thuwal 23955-6900, Saudi Arabia. E-mail: marc.genton@kaust.edu.sa \\
This research was supported by KAUST.}
\end{center}

\baselineskip=14pt \vskip 10mm \centerline{\today} \vskip 15mm

%%%%%%%%%%%%%%%%%%%%%%%%%%%%%%%%%%%%%%%%%%%%%%%%%%%%%%%%%%%%%%%%%%%%%%%%
\begin{center}
{\large{\bf Abstract}}
\end{center}

We propose a new class of extreme-value copulas which are extreme-value limits of conditional normal models. Conditional normal models are generalizations of conditional independence models, where the dependence among observed variables is modeled using one unobserved factor. Conditional on this factor, the distribution of these variables is given by the Gaussian copula. This structure allows one to build flexible and parsimonious models for data with complex dependence structures, such as data with spatial dependence or factor structure. We study the extreme-value limits of these models and show some interesting special cases of the proposed class of copulas. We develop estimation methods for the proposed models and conduct a simulation study to assess the performance of these algorithms. Finally, we apply these copula models to analyze data on monthly wind maxima and stock return minima. 

\baselineskip=14pt

\par\vfill\noindent
\par\medskip\noindent
{\bf Short title}: Conditional Normal Extreme-Value Copulas
\clearpage\pagebreak\newpage \pagenumbering{arabic}
\baselineskip=25pt

\section{Introduction}

Copula models have become more popular in modeling non-Gaussian data. A copula function is a multivariate cumulative distribution function (cdf) with standard uniform $U(0,1)$ marginal cdfs. \cite{Sklar1959} showed that for any $d$-variate continuous cdf, $F_{1,\ldots,d}(z_1,\ldots,z_d)$, with univariate marginal cdfs, $F_1,\ldots,F_d$, there exists a unique copula, $C_{1,\ldots,d}$, such that $F_{1,\ldots,d}(z_1,\ldots,z_d) = C_{1,\ldots,d}\{F_1(z_1),\ldots,F_d(z_d)\}$ for any $z_1,\ldots,z_d$. The converse part of Sklar's theorem implies that the copula can be used to construct a multivariate distribution with given marginal cdfs. This allows greater flexibility when modeling multivariate data.

Many existing copula families, however, are not suitable for modeling multivariate data because they cannot generate flexible dependence structures. For example, most Archimedean copulas have exchangeable dependence \citep{McNeil.Frey.ea2005} and the multivariate Student-$t$ copula is reflection symmetric; asymmetric versions of this copula have been proposed in the literature, but their parameter estimation can be computationally demanding \citep{Yoshiba2018}.

Vine copula models can be used to construct very flexible distributions; see, for example, \cite{Aas.Czado.ea2009} and \cite{Kurowicka.Cooke2006}. These models require the estimation of $O(d^2)$ dependence parameters and might therefore be computationally demanding if $d$ is large. Truncated vines assume independence after conditioning on some variables and have $O(d)$ parameters \citep{Brechmann.Czado.ea2012}. However, vine copula models lack interpretability and might be unable to capture some features of data. One example is data with spatial structure when dependence is weaker with larger distance, and this property is generally not satisfied by vine copula models.

Conditional independence models constitute another class of models in which observed variables are assumed to be independent, conditionally on unobserved (latent) factors. These parsimonious and flexible models can be used for modeling data when all variables can be simultaneously affected by unobserved driving variables, such as financial stock returns \citep{Krupskii.Joe2013,Krupskii.Joe2015b,Oh.Patton2017} and spatial data \citep{Hua.Xia.ea2017}. Extreme-value limits of these models can be used to analyze multivariate extremes data with a factor structure \citep{Lee.Joe.2018}.

However, in many applications these latent factors may not explain all the dependence among the observed variables. To achieve greater flexibility, it is therefore useful to assume some residual dependence after conditioning on the unobserved factors. For example, \cite{Krupskii.Huser.ea2016, Krupskii.Genton2017, Krupskii.Genton2019},  proposed a copula for modeling a non-Gaussian spatial process. In this model, the variables have a multivariate Gaussian distribution after conditioning on an unobserved latent variable which does not depend on a spatial location. %This variable is used to introduce tail dependence for the corresponding copula, and the conditional Gaussian distribution allows one to get weaker dependence with a larger distance. 
\cite{Krupskii.Joe.ea2018} showed that the extreme-value limit of this copula is the H\"{u}sler-Reiss copula \citep{Husler.Reiss1989} which is a flexible model popular in different applications \citep{Aloui.Aissa.ea2014, Bargaoui.Bardossy.2015}.

In this paper, we study extreme-value limits of conditional normal copula models proposed by \cite{Krupskii.Joe2019}. Let $\UU = (U_1,\ldots,U_d)^{\top}$, $U_j \sim U(0,1)$, $j=1,\ldots,d$. Assume that $V_0 \sim U(0,1)$ is the unobserved (latent) variable and that $(\UU|V_0 = v_0) \sim C_N(\cdot;\Sig)$ where $C_N(\cdot;\Sig)$ is the Gaussian (Normal) copula with the correlation matrix $\Sig$. Let $C_{j, 0}$ be the copula linking $U_j$ and $V_0$ and let $C_{j|0}(u_j|v_0) = \p C_{j, 0}(u_j, v_0)/ \p v_0$ be the conditional copula cdf. Throughout this paper, we use small letters for the corresponding copula probability density functions (pdfs). We also assume that all  bivariate copulas $C_{j,0}$ and their densities are strictly positive continuous functions on $(0,1)^2$.

The joint copula cdf and pdf for $\UU$ at $\uu=(u_1,\ldots,u_d)^{\top}$ are
\begin{eqnarray}
C_{\UU}(\uu;\Sig) &=& \int_0^1C_N\{C_{1|0}(u_1| v_0),\ldots,C_{d|0}(u_d| v_0);\Sig\}\df v_0,\nonumber \\
c_{\UU}(\uu;\Sig) &=& \int_0^1c_N\{C_{1|0}(u_1| v_0),\ldots,C_{d|0}(u_d| v_0);\Sig\}\cdot \prod_{j=1}^dc_{j,0}(u_j, v_0)\df v_0.\label{eq1_coppdf}
\end{eqnarray}
Note that in the general case the matrix $\Sig$ can depend on the conditioning value $V_0 = v_0$. 

These models form the general class of models where the variables are assumed to have a multivariate Gaussian distribution after conditioning on an unobserved latent factor.
\cite{Manner.Segers.2011} considered correlation mixtures of elliptical copulas and applied these models to stock market data. The proposed models include a bivariate normal copula with a correlation parameter driven by a latent random process which is a special case of conditional normal copulas.

In this paper, we consider upper tail extreme-value limits of the copula model (\ref{eq1_coppdf}).  The respective limiting extreme-value copula is defined as
\begin{equation}
\label{eq-evcop}
C_{\UU}^{\EV}(u_1, \ldots, u_d) = \exp[- \ell\{-\ln (u_1), \ldots, -\ln (u_d)\}]\,.
\end{equation} 
Here, $\ell(w_1, \ldots, w_d) = V(1/w_1, \ldots, 1/w_d)$ is the stable (upper) tail dependence function and $V(w_1, \ldots, w_d)$ is the exponent function \citep{deHaan.Ferreira.2006}, and
$$
\ell(w_1, \ldots, w_d) = \lim_{u \to 0} \frac{1}{u}\left\{1-C_{\UU}(1 - uw_1, \ldots, 1 - uw_d)\right\}\,,
$$
assuming this limit exists.

We show that these extreme-value limiting copulas are computationally tractable and they can be used for modeling data with complex dependencies, such as multivariate extremes with dynamic dependence, or spatial extremes. Models with a multifactor structure can also be obtained when the multivariate Gaussian distribution has a factor correlation structure, i.e., when $C_N$ is a normal factor copula \citep{Krupskii.Joe2013}.

The rest of this paper is organized as follows. In Section \ref{sec-evlim}
%\ref{sec_1fctnormal} we review dependence properties of conditional normal copulas (\ref{eq1_coppdf}) and 
we define the respective limiting extreme-value copulas, study their properties and consider some special cases with parsimonious dependence. We provide more details on parameter estimation in Section \ref{sec_MLE}, conduct Monte Carlo simulations to assess the performance of the proposed estimators in Section \ref{sec_simul}, and then apply the proposed models to analyze monthly wind maxima and monthly stock returns minima data in Section~\ref{sec_empstudy}. Section \ref{sec_disc} concludes with a discussion.

\section{Extreme-value limits of conditional normal copulas and their properties}
\label{sec-evlim}
%\label{sec_1fctnormal}

\subsection{Extreme-value limits}
\label{subsec-EVlim}

We now consider upper tail extreme-value limits of the copula model (\ref{eq1_coppdf}). Throughout this section, we assume that the linking copula densities $c_{j,0}$, $j=1,\ldots, d$ are continuous functions of their arguments.

%Lower tail limits can be considered analogously. 
Note that the linking copulas $C_{j,0}$, $j=1,\ldots,d$, with upper tail quadrant independence (e.g., if the reflected copula $C^R_j(u,u)=-1+2u +C_{j,0}(1-u,1-u) \sim u^2\ell_j(u)$ as $u \to 0$ where $\ell_j$ is a slowly varying function) are not suitable if $\Sig$ is a non-degenerate matrix as $C_{\UU}$ has no upper tail dependence in this case \citep{Krupskii.Joe2019}.

When using linking copulas $C_{j,0}$, $j=1,\ldots,d$, with intermediate upper tail dependence (e.g., if $C^R_j(u,u) \sim u^{\kappa}\ell_j(u)$ with $1 < \kappa < 2$ as $u \to 0$ where $\ell_j$ is a slowly varying function; see \cite{Hua.Joe2011}), the copula $C_{\UU}$ also has intermediate upper tail dependence in many cases and therefore cannot be used to obtain non-trivial extreme-value limits. For example, if $C_{j,0}$ are the Gaussian copulas with correlation parameters $\rho_j < 1$, then the copula $C_{\UU}$ is the Gaussian copula with the correlation matrix $\Sig^*$ such that $(\Sig^*)_{j,k} = \rho_j\rho_k + \sqrt{(1-\rho_j^2)(1-\rho_k^2)}(\Sig)_{j,k} < 1$, $1 \leq j < k \leq d$. 
The next proposition shows that intermediate upper tail dependence for the copula $C_{\UU}$ can also be obtained when using reflected extreme-value linking copulas $C_{j,0}$. We prove an equivalent result that intermediate lower tail dependence can be obtained when using extreme-value linking copulas.

\begin{prop}  \label{prop3} \rm
Assume that $\Sig$ is non-degenerate and $C_{j,0}(u_1,u_2) = (u_1u_2)^{A_j\{\ln (u_2)/\ln (u_1u_2)\}}$ where $A_j(\cdot): [0,1] \mapsto [1/2,1]$ is a convex function such that $A_j(t) \geq \max(t, 1-t)$, $j=1,\ldots,d$. We also assume that $A_j(t)$ is a continuously differentiable function with $A_j'(t) > -1$ for $t \in (0,0.5)$, $j=1,\ldots,d$. Let the lower tail order of $C_{j,0}$ be $\kappa_j = 2A_j(1/2) > 1$. It implies that $C_{\UU}$ is a copula with intermediate lower tail dependence.
\end{prop}
\emph{Proof:} See Appendix \ref{sec-appx0a}.

%\subsection{Extreme-value limits of conditional normal copulas}
%\label{sec-evlim}

To get non-trivial limits with $C_N$ and non-degenerate $\Sig$, one therefore needs to select %we need to consider the case when the linking copulas $C_{j,0}$, $j=1,\ldots,d$, have upper tail dependence because the copula $C_{\UU}$ has upper tail dependence in this case.
the linking copulas $C_{j,0}$, $j=1,\ldots,d$, with upper tail dependence as the copula $C_{\UU}$ has upper tail dependence in this case. 

\begin{prop}  \label{prop4} \rm Assume that $b_{j|0}(w_j|w_0) = 1-\lim_{u\to 0}C_{j|0}(1-uw_j|1-uw_0)$ and $b_{0|j}(w_0|w_j) = 1-\lim_{u\to 0}C_{0|j}(1-uw_0|1-uw_j)$ where $b_{j|0}(\cdot|w_0)$ is a proper distribution function for any $w_0 > 0$ and $b_{0|j}(\cdot|w_j)$ is a proper distribution function for any $w_j > 0$. It follows that the stable tail dependence function of $C_{\UU}^{\EV}$ is
	$$\ell(w_1,\ldots,w_d) = \int_0^{\infty}\left[1-C_N\{1-b_{1|0}(w_1|w_0),\ldots,1-b_{d|0}(w_d|w_0); \Sig\} \right]\df w_0, \  w_1,\ldots,w_d > 0.$$
\end{prop}
\emph{Proof:} See Appendix \ref{sec-appx0b1}.

Many linking copulas with upper tail dependence satisfy the assumption of this proposition, including the Gumbel and reflected Clayton copulas; see \cite{Joe2014} for more details on bivariate copula families and their dependence properties. With continuously differentiable tail functions $b_{j|0}(w_j|w_0)$, $j=1, \ldots,d$, the stable tail dependence function in Proposition \ref{prop4} is also a continuously differentiable function and therefore the resulting extreme-value copula is an absolutely continuous copula if $\Sig$ is not a singular matrix. If $C_N$ is the comonotonicity copula (i.e., $\Sig = \boldsymbol{J_d}$ is the matrix of ones), then the formula for $\ell(w_1,\ldots,w_d)$ simplifies to
\begin{equation*}
%\label{eq-rho1-case1}
\ell(w_1,\ldots,w_d)= \int_0^{\infty}\max_j \{b_{j|0}(w_j|w_0)\}\df w_0.
\end{equation*}

One interesting class of limiting extreme-value copulas %with singular components 
arises when $C_{N}$ is the comonotonicity copula and $c_{j,0}$, $j=1,\ldots,d$, are continuous functions on $(0,1]^2$. 

\begin{prop}  \label{prop5} \rm Assume that $\Sig = \boldsymbol{J_d}$ is a matrix of ones and $c_{j,0}(u_j, u_0)$ are continuous functions on $(0,1]^2$ and positive on $(0,1)^2$, $j=1,\ldots,d$. The stable tail dependence function of $C_{\UU}^{\EV}$ is
	\begin{equation} \label{eq-rho1-case2}
	\ell(\ww) = \ell(w_1,\ldots,w_d) = \int_0^1 \max_j\{w_jc_{j,0}(1,v_0)\} \df v_0, \quad w_1,\ldots,w_d > 0.
	\end{equation}
\end{prop}
\emph{Proof:} See Appendix \ref{sec-appx0b2}. 

Many copulas with upper tail quadrant independence satisfy the conditions of Proposition \ref{prop5}, including the Clayton and Frank copulas. 

\subsection{Properties and special cases}
\label{subsec-example}

In this section we consider some special cases and study the properties of bivariate margins of the extreme-value limiting copulas $C_{\UU}^{\EV}$. In the next two examples we use observed and latent variables that do not have a uniform distribution; however, the respective copula is invariant with respect to monotone transformations of the random variables.

\medskip
\textbf{Example 1:} The H\"{u}sler-Reiss copula \citep{Husler.Reiss1989} can be obtained as an extreme-value limit of the convolution of exponential and multivariate normal distributions \citep{Krupskii.Joe.ea2018}. In this model, the copula $C_{\UU}$ corresponds to the joint distribution of $\WW = (W_1,\ldots,W_d)^{\top}$, where $W_j = Z_j + \alpha_j V_0$, $\ZZ = (Z_1,\ldots,Z_d)^{\top}$ has a multivariate normal distribution with $N(0,1)$ marginals and the correlation matrix $\Sig$ and where $\ZZ$ is independent of $V_0$. Here, $C_{j,0}$ links $W_j = Z_j + \alpha_j V_0$ and $V_0 \sim \mathrm{Exp}(1)$, and one can show that the tail function of this copula is $b_{j|0}(w_j|w_0) = \Phi\left(\alpha_j\ln\frac{w_j}{w_0}-\frac{1}{2\alpha_j}\right)$, $j=1,\ldots,d$. \cite{Castro.Huser.2020} developed the local likelihood approach to estimate parameters of this model for non-stationary data with spatial dependence and applied it to U.S. precipitation extremes data.

\medskip
\textbf{Example 2:} An extremal $t$ copula \citep{Demarta.McNeil.2005} can be obtained as an extreme-value limit of the $t$ copula with $\nu > 0$ degrees of freedom \citep{Nikoloulopoulos.Joe.ea2009}.  In this model, the copula $C_{\UU}$ corresponds to the joint distribution of $\WW = (W_1,\ldots,W_d)^{\top}$, where $W_j =  V_0 Z_j$, $\ZZ = (Z_1,\ldots,Z_d)^{\top}$ has the multivariate normal distribution with $N(0,1)$ marginals and the correlation matrix $\Sig$ and where $\ZZ$ is independent of $V_0$. Here, $C_{j,0}$ links $W_j = V_0 Z_j$ and $V_0  = (Y/\nu)^{-1/2}$ with $Y \sim \chi^2(\nu)$.

Now consider the (1,2) marginal of $C_{\UU}^{\EV}$ which we denote by $C_{12}^{\EV}$. Let $\varrho = \Sig_{1,2}$. The following properties hold: 
\begin{enumerate}
	\item If $C_{1,0} = C_{2,0}$, then $C_{12}^{\EV}$ is a permutation symmetric copula.  The opposite is not necessarily true;
	\item Under the assumptions of Proposition \ref{prop4}, if $b_{j|0}(w_j|w_0)$ is a strictly increasing function of $w_j$ for $0 < w_0 < \infty$, $j=1,2$, and $|\varrho| < 1$, then the support of the density $c_{1,2}^{\EV}$ is the unit square $[0,1]^2$; 
	\item  Under the assumptions of Proposition \ref{prop4}, if $b_{j|0}(w_j|w_0)$ is a continuously differentiable function on $(0,1)^2$, and $|\varrho| < 1$, then $C_{\UU}^{\EV}$ is absolutely continuous;
	\item Under the assumptions of Proposition \ref{prop5}, the support of the density $c_{1,2}^{\EV}$ is $$\{(u_1, u_2): u_1^{\xi_L} < u_2 < u_1^{\xi_U}\}, \quad \xi_L = \max_{v_0 \in (0,1)} \frac{c_{2,0}(1,v_0)}{c_{1,0}(1,v_0)}, \,\xi_U = \min_{v_0 \in (0,1)} \frac{c_{2,0}(1,v_0)}{c_{1,0}(1,v_0)}\,.$$
\end{enumerate}

Note that the H\"{u}sler-Reiss copula is permutation symmetric for $d=2$ despite $C_{1,0} \neq C_{2,0}$ in the general case. This copula is not permutation symmetric for $d > 2$. Property 2 holds since $C_{12}^{\EV}(u_1, u_2)$ is a strictly increasing function of $0< u_1, u_2 < 1$ in this case. This property holds for many copulas with upper tail dependence including the reflected Clayton and Gumbel copulas.  

Under mild regularity conditions, property 3 also holds if $\varrho = 1$ and $C_{1,0} \neq C_{2,0}$. For example, if $c_{1,0}$ and $c_{2,0}$ are twice continuously differentiable functions on $(0,1)^2$ and there exists  $0<v^*<1$ such that $w_1 c_{1,0}(v,1) > w_2 c_{2,0}(v,1)$ if and only if $v > v^*$ (satisfied for many copulas with upper tail quadrant independence including the Farlie-Gumbel-Morgenstern and Clayton copulas), then 
$$
\ell(w_1, w_2) = w_2 C_{2|0}(v^*|1) + w_1 - w_1C_{1|0}(v^*|1),
$$
where $v^* = v^*(w_1, w_2)$ is a twice continuously differentiable function of $w_1$ and $w_2$ by the implicit function theorem. It implies that (\ref{eq-rho1-case2}) is a twice continuously differentiable function of $w_1$ and $w_2$. 

Assuming $w_1 \neq w_2$, property 4 holds since $C_{1,2}^{\EV}(u_1, u_2) = u_1$ 
(equivalently, $\ell(w_1, w_2) = w_1$) if and only if $w_1c_{1,0}(1,v) \geq w_2c_{2,0}(1,v)$ for any $v \in (0,1)$. 
Similarly, $C_{1,2}^{\EV}(u_1, u_2) = u_2$ if and only if $w_1c_{1,0}(1,v) \leq w_2c_{2,0}(1,v)$ for any $v \in (0,1)$. 
%and $\ell(w_1, w_2) = w_2$ if and only if $w_1c_{1,0}(1,v) \leq w_2c_{2,0}(1,v)$ for $w_1 \neq w_2$ and $0 < v < 1$.

%It follows that the support of the corresponding limiting copulas is a subset of $[0,1]^d$. For example, if $w_1$ is such that $w_1\min_{0<v_0<1}\{c_{j,0}(1,v_0)\} > \max_{0<v_0<1, k \neq 1} \{w_kc_{k,0}(1,v_0)\}$, then $V(w_1,\ldots,w_d) = w_1$ and hence the density in the neighborhood of the corner $(1, 0, \ldots, 0)^{\top}$ is zero. This class of copulas can therefore be used for modeling multivariate extremes with very strong dependence (close to comonotonic dependence).  

\medskip
\textbf{Example 3:} Let $d=2$ and $C_{1,0}$ and $C_{2,0}$ are the Farlie-Gumbel-Morgenstern copulas with parameters $\theta_1 = 0.5$ and $\theta_2=-0.5$, respectively. One can find that 
$$
\ell(w_1, w_2) = \begin{cases}
\frac{9w_1^2 + 2w_1w_2+9w_2^2}{8(w_1+w_2)}\,, & \text{ if } \frac{1}{3} < \frac{w_1}{w_2} < 3,\\
w_1, & \text{ if } \frac{w_1}{w_2} > 3,\\
w_2, & \text{ if } \frac{w_1}{w_2} < \frac{1}{3}\,.\\
\end{cases}
$$
This implies that the limiting copula density is positive only if $u_2^3 < u_1 < u_2^{1/3}$. Note that $$\xi_L = \max_{v_0 \in (0,1)}\frac{1-\theta_2(1-2v_0)}{1-\theta_1(1-2v_0)} = \max_{v_0 \in (0,1)}\frac{1.5-v_0}{0.5 + v_0} = 3\,$$
and similarly $\xi_U = 1/3$ for this copula.

It is seen that the stable tail dependence function is a continuously differentiable function with respect to $w_1$ and $w_2$ and hence the conditional limiting copula is absolutely continuous. The copula density is positive for  $u_2^3 < u_1 < u_2^{1/3}$.

\subsection{Parsimonious dependence structures}
\label{subsec-parsim}

Different types of dependencies can be generated by the extreme-value limits of $C_{\UU}$, depending on the choice of the linking copulas, $C_{j,0}$, $j=1,\ldots,d$, and the correlation matrix $\Sig$. We now consider some special cases resulting in parsimonious dependence structures for the extreme-value limit of the copula $C_{\UU}$ with $O(d)$ dependence parameters.
\bigskip

\subsubsection{Copulas with spatial dependence}
%\medskip
\label{subsec-parsim-spat}

If $\Sig$ is a spatial correlation matrix, then the limiting copula can be useful for modeling spatial extremes. Different covariance functions can be used to construct the covariance matrix with a spatial structure; see \cite{Gneiting.Genton.ea2007} for an overview of covariance functions. 
One should select the same linking copulas $C_{1,0}= C_{2,0} = \cdots = C_{d,0}$ to ensure that the $(j,k)$ margin converges to the comonotonicity copula when $\Sig_{jk} \to 1$ for $1 \leq j < k \leq d$. Otherwise, one gets the limiting distribution as described in Proposition \ref{prop5}, which is not a comonotonicity copula in the general case.  

Assuming the same linking copulas are used, one can control the rate of decay of tail dependence as a function of distance or time lag by selecting an appropriate spatial correlation matrix, as the next proposition shows.

\begin{prop}  \label{prop6} \rm 
    Consider the (1,2)-margin of the copula $C_{\UU}^{\EV}$ with $C_{1|0} = C_{2|0}$ defined in (\ref{eq-evcop}). Assume that $b_{1|0}(1|w) \sim w^{-\theta}\tilde\ell(w)$ as $w \to \infty$ for $\theta > 1$, where $\tilde\ell(w)$ is a slowly varying function. Under the assumptions of Proposition \ref{prop4}, $$\ell(1,1) = 1 + \left(\frac{1-\rho}{\pi}\right)^{1/2}\int_0^{\infty} \phi[\Phi^{-1}\{b_{1|0}(1|w_0)\}] \df w_0 + O((1-\rho)^{3/2}), \quad \text{ as } \rho \to 1,$$
    where $\Phi(\cdot)$, $\phi(\cdot)$ are the standard normal cdf and pdf, respectively, and $\rho = \Sig_{1,2}$.
\end{prop}
\emph{Proof:} See Appendix \ref{sec-appx0}. 

Assume that $\rho(r) = 1 - \eta -  Cr^{\alpha} + o(r^{\alpha})$ as $r \to 0$ for some constants $C, \alpha > 0$, where $0 < \eta < 1$ is a nugget effect\footnote[3]{The nugget effect is used to describe the variability of measurements that are closely spaced (e.g., due to measurement errors)} and $r$ is the distance between two locations. Proposition \ref{prop6} implies that the upper tail dependence coefficient corresponding to the (1,2)-margin of the copula $C_{\UU}^{\EV}$ is $$\lambda_U(r) = 2 - \ell(1,1) = 1 - \left(\frac{\eta+Cr^{\alpha}}{\pi}\right)^{1/2}\int_0^{\infty} \phi[\Phi^{-1}\{b_{1|0}(1|w_0)\}] \df w_0  + o(r^{\alpha/2}).$$
A correlation function $\rho$ with a smaller $\alpha$ can therefore be selected to obtain a faster rate of decay of the tail dependence as a function of distance $r$ when it is close to zero.

To model multivariate spatial extremes, different linking copulas (they can be from different parametric families or from the same parametric family but have different parameters) can be used for different variables together with the cross-correlation matrix $\Sig$; see \cite{Genton.Kleiber2015}  for a review on cross-covariance models. For example, to model bivariate spatial extremes, one can select $C_{1,0}= C_{2,0} = \cdots = C_{m,0}$ for the first variable and $C_{m+1,0}= C_{m+2,0} = \cdots = C_{d,0}$ for the second variable, where $d=2m$ and $m$ is the number of spatial locations.

To model spatial isotropy, one can select an isotropic correlation matrix $\Sig$, and for non-stationary data, one can select a non-stationary correlation matrix $\Sig$.

%\bigskip
\subsubsection{Copulas with factor structure and dynamic dependence}
%\medskip

The limiting extreme-value copula with a one factor structure can be obtained with $\Sig=\boldsymbol{I_d}$; this class of models was studied by \cite{Lee.Joe.2018}, who considered continuous copulas. Copulas with singular components can be obtained by the discontinuous functions $b_{j|0}(w_j|w_0)$, $j=1,\ldots,d$. These copulas can be used to model the lifetime of system components that may fail simultaneously, with a positive probability. One example is Marshall-Olkin copulas with a factor structure \citep{Krupskii.Genton2018}. One application of such copulas is to model the times-to-default for components of a credit portfolio. A correlation matrix $\Sig$ with a one-factor structure can be used to obtain two-factor structure models, and bi-factor models can be obtained if $\Sig$ is a block-diagonal matrix. In all these cases, a one-dimensional integration is required to compute the copula density. \cite{Lee.Joe.2018} have provided some details about extreme-value copulas with two-factor structures; however, their approach requires the computation of two-dimensional integrals and is therefore not feasible in very high dimensions. 

%\bigskip
%\subsection{Dynamic dependence}
%\medskip

Dynamic dependence can be  modeled by selecting a matrix $\Sig$ with a simple parsimonious structure with time-varying correlations and linking copulas $C_{j,0}$, $j=1,\ldots,d$, that do not change over time. The resulting copula in (\ref{eq1_coppdf}) can be used to model data with dynamic extremes, such as stock returns' monthly maxima or minima, or time series extremes.

%\medskip
%\textbf{Example 4:} Let $t=1,\ldots,T$, and we assume $\Sig = \Sig(t)$ with time-varying correlations: $(\Sig_t)_{j,k}= \rho(t)\rho_{j}\rho_{k}$, $1 \leq j \leq k \leq d$, where $0 \leq \rho_1, \ldots, \rho_d \leq 1$ and 
%$$
%\rho(t) = \frac{1}{1 + \exp\{\eta(t)\}}, \quad \eta(t) = \beta_0 + \sum_{k=1}^K \beta_k V_k(t),$$ 
%where $V_1(t), \ldots, V_K(t)$ are time-dependent  external variables used to model nonstationarity in time series. Here, $\Sig$ is a correlation matrix of $\ZZ = (W_{1}, \ldots, W_{d})^{\top}$ where
%\begin{equation}
%\label{eq-fctmodel}
%W_{i} = \{\rho(t)\}^{1/2}\rho_{i} Z_0  + \{1-\rho(t)\rho_i^2\}^{1/2}Z_{i}, \quad i = 1, \ldots, d,
%\end{equation}
%and $Z_0, Z_1, \ldots, Z_d \sim_{\text{i.i.d.}} N(0,1)$. This correlation structure can be combined with linking copulas $C_{j,0}$ that have lower/upper tail dependence to model minima/maxima time series data. \cite{Krupskii.Joe2019} used the conditional normal copula $C_{\UU}$ with correlation structure (\ref{eq-fctmodel}) and with BB1 copulas $C_{j,0}$ to model dynamic dependence in European bonds, credit default swaps, and US stock returns data.   

\subsection{Parameter identifiability}

Under the assumptions of Proposition \ref{prop6}, the asymptotic behavior of the upper tail dependence coefficient for each pair of variables depends on the parameters of the correlation matrix $\Sig$, so these parameters and the linking copula parameters can be uniquely identified for copulas with a spatial dependence structure defined in Section \ref{subsec-parsim-spat} provided that the data set includes locations at small distances. It is difficult to formally check that parameters are identifiable for the proposed class of models in the general case. In Section \ref{sec_simul} we show that accurate parameter estimates %proper convergence of the parameter estimates to the true values 
can be obtained for $C_{\UU}^{\EV}$ with a factor structure and the reflected Clayton linking copulas. %, so this model is also numerically identifiable. 
We obtained similar results with the Gumbel linking copulas.

\medskip
\textbf{Example 4:} If $b_{j|0}(w_j|w_0) = \Phi\left(\alpha_j \ln \frac{w_j}{w_0} - \frac{1}{2\alpha_j}\right)$ with $\alpha_j > 0$, $j=1,\ldots, d$, then $C_{\UU}^{\EV}$ is the H\"{u}sler-Reiss copula with parameters $$\eta_{ij} = \frac{1}{\alpha_i\alpha_j}(\alpha_i^2 + \alpha_j^2 - 2\Sig_{i,j}\alpha_i\alpha_j)^{1/2}\,, \quad i \neq j.$$
There are $d(d-1)/2$ functionally independent parameters that can be identified from bivariate marginals. If $\Sig_{i,j} = \rho_i \rho_j$, $i \neq j$ for $-1 < \rho_i < 1$, $i, j = 1, \ldots, d$, then the model has $2d$ independent parameters and these can be identified if $2d \leq d(d-1)/2$, i.e., $d \geq 5$.

Parameters of copula models with the stable tail dependence function as given in (\ref{eq-rho1-case2}) might be non-identifiable for some linking copulas. In particular, parameters are not identifiable if the Clayton linking copulas are used. Even though it can be shown that parameters are identifiable for some other linking copulas such as the Farlie-Gumbel-Morgenstern copulas, numerical estimates for these models are not accurate. The reason is that different sets of parameters result in very similar copulas. Some parameters can be fixed to obtain consistent estimates of the remaining parameters for this class of models; we provide more details in Sections \ref{sec_MLE} and \ref{sec_simul}.

%Numerical experiments show that this model is nearly non-identifiable with other linking copulas including the Farlie-Gumbel-Morgenstern copulas, so some parameters can be fixed to obtain consistent estimates of the remaining parameters; we provide more details in Sections \ref{sec_MLE} and \ref{sec_simul}.

\section{Parameter Estimation}
\label{sec_MLE}

In this section, we show how the parameters of $C_{\UU}^{\EV}$ in (\ref{eq-evcop}) can be estimated. We assume that $\tht_j$ is a vector of parameters for linking copulas $C_{j,0}$, $j=1,\ldots, d$, and $\tht_{\Sig}$ is a vector of parameters for the correlation matrix $\Sig$. %Table shows the root mean squared errors (RMSE) for the two estimation methods for the simulated data sets of size $n=200$ and $n=1000$. The results reported in this table are based on $N=500$ simulations. 

\subsection{Composite likelihood inference for $C_{\UU}^{\EV}$ with non-degenerate~$\Sig$}
\label{subsec-mle-est}

Here we show how to obtain parameter estimates for the copula $C_{\UU}^{\EV}$ using the maximum likelihood approach. We assume that the conditions of Proposition \ref{prop4} are satisfied. As likelihood inference is not feasible for extreme-value copulas in high dimensions, we will use the pairwise likelihood approach to estimate these parameters; see, for example, \cite{Lindsay.1998}.

Let $\{\uu_i = (u_{1i},\ldots,u_{di})^{\top}\}_{i=1}^n$ be a sample of size $n$ from $C_{\UU}^{\EV}$. In applications, models of univariate data can be fitted first, and the probability integral transform can be applied to convert the original data to uniform $U(0,1)$ data. Alternatively, uniform scores data can be obtained using rank transforms \citep{Genest.Ghoudi.ea1995}. Under mild regularity conditions, the estimators obtained using the two-step approach are consistent and asymptotically normal \citep{Andersen2004}. In particular, the two-step estimators are a special case of the generalized method of moments estimators and they are consistent and asymptotically normal if the model parameters are identifiable and some  regularity conditions hold; see Chapter 36, Theorems 2.6, 3.4 and 6.1 in \cite{Newey.McFadden1994}.
We do not provide details in this paper, and a more detailed study of asymptotic properties of the model estimators is a topic of future research.

The pairwise log-likelihood is
$$
\ell(\uu_1,\ldots,\uu_n; \tht) = \sum_{i=1}^n \sum_{1 \leq j < k \leq d} \ln c^{\EV}_{j,k}(u_{ji}, u_{jk}; \tht_j, \tht_k, \tht_{\Sig}), \quad \tht = (\tht_1^{\top}, \ldots, \tht_d^{\top}, \tht_{\Sig}^{\top})^{\top}.
$$
The copula pdf of the $(j,k)$-margin is
$$
c^{\EV}_{j,k}(u_j, u_k) = C^{\EV}_{j,k}(u_j,u_k)\left\{\frac{\p V_{j,k}(w_j, w_k)}{\p w_j}\cdot\frac{\p V_{j,k}(w_j, w_k)}{\p w_k} - \frac{\p^2 V_{j,k}(w_j, w_k)}{\p w_j \p w_k}\right\}\,,
$$
where $w_j = -\ln u_j, w_k = -\ln u_k$, $C_{j,k}^{\EV}(u_j, u_k) = \exp\{-V_{j,k}(w_j,w_k)\}$ and
{\small \begin{eqnarray*}
V_{j,k}(w_j, w_k) &=& \int_0^{\infty}\left[1-C_N\{1-b_{j|0}(w_j|w_0),1-b_{k|0}(w_k|w_0); \rho_{jk}\}\right] \df w_0, \quad \rho_{jk} = \Sig_{j,k},\\
\frac{\p V_{j,k}(w_j, w_k)}{\p w_j} &=& \int_0^{\infty} C_N(1-b_{k|0}(w_k|w_0)| 1 - b_{j|0}(w_j|w_0); \rho_{j,k}) b_{j,0}(w_j, w_0)\df w_0, \\
\frac{\p V_{j,k}(w_j, w_k)}{\p w_k} &=& \int_0^{\infty} C_N(1-b_{j|0}(w_j|w_0)| 1 - b_{k|0}(w_k|w_0); \rho_{j,k}) b_{k,0}(w_k, w_0)\df w_0, \\
\frac{\p^2 V_{j,k}(w_j, w_k)}{\p w_k \p w_j} &=& -\int_0^{\infty}c_N\{1-b_{j|0}(w_j|w_0),1-b_{k|0}(w_k|w_0); \rho_{jk}\}b_{j,0}(w_j, w_0) b_{k,0}(w_k, w_0) \df w_0,
\end{eqnarray*} }
where $b_{j,0}(w_j, w_0) = \p b_{j|0}(w_j | w_0)/\p w_j$, $b_{k,0}(w_k, w_0) = \p b_{k|0}(w_k | w_0)/\p w_k$.

\vspace{2mm}
Numerical integration is required to compute $V_{j,k}(w_j, w_k)$ and its derivatives. The integrand is a slowly decaying function of $w_0$, so some changes in variables are required to make the computation more efficient. More details about computing $V_{j,k}(w_j,w_k)$ are provided in Appendix \ref{sec-appx1}. A quasi-Newton-Raphson method can then be used to estimate a vector of parameters $\tht$ for the copula $C_{\UU}^{\EV}$.

\subsection{Inference for $C_{\UU}^{\EV}$ with degenerate $\Sig$}
\label{subsec-mm-est}

In this section, we assume that the conditions of Proposition \ref{prop5} are satisfied for $C_{\UU}^{\EV}$. The pairwise likelihood approach is not feasible in this case, as the support of the $(j,k)$ marginal copula depends on $\tht_j$ and $\tht_k$, and it can be very difficult to identify the parameter values $\tht$ where the copula density is positive if the dimension $d$ is large. 

Instead, we can use non-parametric estimates of the upper tail dependence coefficient $\lambda_{j,k} = \lambda_{j,k}(\tht_j, \tht_k) = 2-V_{j,k}(1,1)$ for $1 \leq j < k \leq d$ \citep{Ferreira2013} and define
\begin{equation}
\label{eq-argmin}
\widehat\tht = \text{argmin} \sum_{1 \leq j < k \leq d} \{\widehat\lambda_{j,k} - \lambda_{j,k}(\tht_j, \tht_k)\}^2, 
\end{equation}
where $\widehat\lambda_{j,k}$ and $\lambda_{j,k}(\tht_j, \tht_k)$ are nonparametric and model-based estimates, respectively. Under mild regularity conditions that are satisfied for many linking copulas, the resulting estimator $\widehat\tht$ is a minimum distance estimator and it is consistent and asymptotically normal if the nonparametric estimators of tail dependence coefficients are consistent and asymptotically normal and the model parameters are uniquely identifiable \citep{Millar.1984}.

If the model parameters are not identifiable, some of them can be fixed and consistent estimates of the remaining parameters can be obtained by minimizing the objective function in (\ref{eq-argmin}). In Appendix \ref{sec-appx2}, we provide model details, including the explicit formula for $V_{j,k}(1,1)$ as a function of $\tht_j$ and $\tht_k$, for $C_{\UU}^{\EV}$ with Clayton linking copulas $C_{j,0}$, $j=1, \ldots, d$. Similarly, one can use different linking copulas with upper tail quadrant independence, such as the Frank copula. 
    
\medskip    
\emph{Remark:} This method can also be used to estimate the parameters of $C_{\UU}^{\EV}$ with a non-degenerate $\Sig$. The computation of $V_{j,k}(1,1)$ for different pairs $(j,k)$ is much easier with this family of copulas, and the respective estimates can be used as starting values for the quasi-Newton-Raphson algorithm discussed in the previous section. We provide more detail in the next section.

\section{Monte Carlo Simulations}
\label{sec_simul}

In this section, we conduct some simulation studies to check the performance of the parameter estimation methods discussed in Section \ref{sec_MLE}. In applications, the data are often observed at a higher frequency (e.g., daily data), and the data can be aggregated to block maxima or minima (e.g., monthly or yearly maxima). We therefore simulate data from the conditional normal copula (\ref{eq1_coppdf}), and then use block maxima of size $n_B = 5, 10, 25, \infty$. We transform the univariate marginals of the block maxima data to uniform scores using nonparametric ranks and estimate copula parameters as described in the previous section.    

For the fast estimation method introduced in Section \ref{subsec-mm-est}, we use a nonparametric estimator of the upper tail dependence coefficient proposed by \cite{Ferreira2013}:
\begin{equation}
\label{eq-lambda}
\widehat\lambda = 3 - \left[1-\frac{1}{n} \sum_{i=1}^n \max(\widehat U_{1i}, \widehat U_{2i})\right]^{-1},
\end{equation}
where $\{(\widehat U_{1i}, \widehat U_{2i})\}_{i=1}^n$ is a sample from a bivariate extreme-value distribution transformed to uniform scores using ranks. This  estimator is consistent and asymptotically normal for extreme-value distributions, and it also gives fairly accurate results for block maxima data.

\subsection{Simulation 1: $\Sig$ has an autoregressive structure, $d=10$} 

We simulate a data set from the model (\ref{eq1_coppdf}) %extreme-value copula (\ref{eq-evcop})
with the reflected Clayton linking copulas that have upper tail dependence $C_{j,0}$, $j=1,\ldots,d$ and $d=10$. We assume that the vector of linking copula parameters is $\tht = (1,1,1,2.5,2.5,2.5,2.5,1.5,1.5,1.5)^{\top}$. We also assume that the correlation matrix has an autoregressive structure with $\Sig_{j,k} = \rho^{|j-k|}$ and $\rho = 0.5$. To estimate the parameters $\tht$ and $\rho$, we first use the fast approach from Section \ref{subsec-mm-est} (method 1), and we use the respective estimates as starting values for the pairwise likelihood approach, as shown in Section \ref{subsec-mle-est} (method 2). 

Table \ref{tab1} shows the bias and root mean squared errors (RMSE) of the parameter estimates $\widehat\tht$ and $\widehat\rho$, obtained by the two methods mentioned above for 500 samples of size $n=60$ and $n=200$. With the block size $n_B = 5$, there is a small bias and the RMSEs of the estimates are larger, especially for large values of the parameters. With $n_B = 25$ and $n_B = \infty$, the bias is almost zero and the parameters estimates have very similar RMSEs. This implies that the block size $n_B=25$ is large enough to obtain accurate estimates.     % which corresponds to 5 and 16 years of monthly maxima.

Table \ref{tab1} shows that the fast method produces accurate estimates; the RMSEs of estimates of $\tht$ and $\rho$ are 5--10\% and 20--25\% higher using this method than with the pairwise likelihood approach. The estimates are less accurate for linking copulas with stronger dependence. 

\begin{table}[h!]
	\caption{{\footnotesize RMSEs and biases (multiplied by 100) of the parameter estimates, $\widehat\tht$ and $\widehat\rho$, obtained by method 1 (first line) and method 2 (second line). We used 500 samples of size $n=60$ and $n=200$ with reflected Clayton copulas and $\tht_{1:3}=1, \tht_{4:7} = 2.5, \tht_{8:10}=1.5$ and $\Sig_{j,k} = \rho^{|j-k|}$ with $\rho=0.5$.  }}
	\label{tab1}
	\begin{center}
		{\small\begin{tabular}{l|cccc|cccc|cccc|cccc}
			\hline
			 block size & 5 & 10 & 25 & $\infty$ & 5 & 10 & 25 & $\infty$ & 5 & 10 & 25 & $\infty$ & 5 & 10 & 25 & $\infty$ \\
			 \hline
			 \hline
			 & \multicolumn{4}{c|}{RMSE of $\widehat\tht_{1:3}$} & \multicolumn{4}{c|}{RMSE of $\widehat\tht_{4:7}$} & \multicolumn{4}{c|}{RMSE of $\widehat\tht_{8:10}$} & \multicolumn{4}{c}{RMSE of $\widehat\rho$}\\
			\hline
			\multirow{2}*{$n=60$} &25&26&24&24&84&77&76&78&44&35&36&38&13&12&11&11\\
			 &24&26&23&24&81&75&74&74&42&33&34&36&12&10&9&10\\
			%$n=60$ &24&26&23&24&81&75&74&74&42&33&34&36&12&10&9.4&9.8\\
			\hline
			\multirow{2}*{$n=200$} &14&13&13&13&51&39&37&36&25&21&19&19&9&7&6&6\\
			 &13&12&13&13&46&36&33&32&23&20&18&18&7&5&5&5\\
			%$n=200$ &14&13&13&13&51&39&37&36&25&21&19&19&8.6&6.6&5.8&6.0\\
			%$n=200$ &13&12&13&13&46&36&33&32&23&20&18&18&6.6&4.9&4.7&4.9\\
			\hline
			\hline
			& \multicolumn{4}{c|}{bias of $\widehat\tht_{1:3}$} & \multicolumn{4}{c|}{bias of $\widehat\tht_{4:7}$} & \multicolumn{4}{c|}{bias of $\widehat\tht_{8:10}$} & \multicolumn{4}{c}{bias of $\widehat\rho$}\\
			\hline
			%$n=60$ &5.8&1.0&0.0&1.0&31&12&0.0&4.8&15&7.3&0.5&2.1&7.2&3.3&1.7&0.3\\
			%$n=60$ &4.3&0.0&0.6&1.3&27&11&1.2&3.5&14&6.6&1.4&0.6&5.1&1.7&0.0&0.6\\
			\multirow{2}*{$n=60$} &$-6$&$-1$&0&1&$-31$&$-12$&~~0&5&$-15$&$-7$&$-1$&2&7&3&2&~~0\\
			&$-4$&~~0&1&1&$-27$&$-11$&$-1$&4&$-14$&$-7$&$-1$&1&5&2&0&$-1$\\
			\hline
			%$n=200$ &6.1&2.6&1.0&&38&19&7.8&&18&9.5&3.8&&6.4&3.3&1.2&0\\
			%$n=200$ &4.6&2.2&0.6&&34&18&8.6&&15&8.1&3.7&&4.3&1.7&0.2&0\\
			\multirow{2}*{$n=200$} &$-6$&$-3$&$-1$&0&$-38$&$-19$&$-8$&3&$-18$&$-10$&$-4$&~~0&6&3&1&~~0\\
			&$-5$&$-2$&$-1$&0&$-34$&$-18$&$-9$&2&$-15$&$-8$&$-4$&$-1$&4&2&0&~~0\\
			\hline			
		\end{tabular}}
	\end{center}
\end{table}

\subsection{Simulation 2: $\Sig$ has an autoregressive structure, $d=20$} 

We simulate a data set from the same copula as in Simulation 1 but with $d=20$, the vector of reflected Clayton linking copula parameters $\tht = (0.8,0.8,0.8,0.8,0.8,1.2,1.2,$ $1.2,1.2,1.2,1.6,1.6,1.6,1.6,1.6,2,2,2,2,2)^{\top}$, and $\rho = 0.7$. Again, we use methods 1 and 2 to estimate the copula parameters and compute the RMSEs based on these estimates; Table~\ref{tab2} shows the RMSEs of the estimates obtained using these two methods for 500 samples of sizes $n=60$ and $n=200$.

Again, the results are very similar with $n_B = 25$ and $n_B = \infty$, and the RMSEs of estimates of $\tht$ and $\rho$ are 5--15\% and 25\% higher with the fast method than with the pairwise likelihood approach. The fast method is less accurate if the dependence is stronger. Both methods yield more accurate estimates of the correlation parameter $\rho$ than those used in Simulation 1, so including more temporal or spatial data  locations can improve estimates of the correlation matrix $\Sig$.  

\begin{table}[h!]
	\caption{{\footnotesize RMSEs and biases (multiplied by 100) of the parameter estimates, $\widehat\tht$ and $\widehat\rho$, obtained by method 1 (first line) and method 2 (second line). We used 500 samples of size $n=60$ and $n=200$ with reflected Clayton linking copulas and $\tht_{1:5} = 0.8, \tht_{6:10} = 1.2, \tht_{11:15} = 1.6, \tht_{16:20} = 2$ and $\Sig_{j,k} = \rho^{|j-k|}$ with $\rho=0.7$. }}
	\label{tab2}
	\begin{center}
		%\begin{tabular}{lccccccccccc}
		%	\hline
		%	sample size & \multicolumn{10}{c}{RMSE of $\widehat\tht$} & RMSE of $\widehat\rho$\\
		%	\hline
		%	\multirow{2}*{$n=200$, method 1} &10&11&11&11&12&17&16&16&18&17&\multirow{2}*{3.1}\\
		%	&22&20&21&22&23&29&26&28&31&37&\\
		%	\hline
		%	\multirow{2}*{$n=200$, method 2} &10&11&10&10&11&16&15&16&17&16&\multirow{2}*{2.5}\\
		%	&21&19&19&20&19&27&24&25&27&34&\\
		%	\hline
		%	\multirow{2}*{$n=1000$, method 1} &4.9&4.8&5.0&4.5&4.9&7.1&6.9&6.5&6.5&6.2&\multirow{2}*{1.4}\\
		%	&9.1&9.8&9.8&10&9.6&13&13&12&13&12&\\
		%	\hline
		%	\multirow{2}*{$n=1000$, method 2} &5.1&4.7&4.8&4.6&4.9&6.8&6.9&6.4&6.2&6.0&\multirow{2}*{1.1}\\
		%	&8.9&8.8&8.7&9.0&8.6&11&11&10&11&10&\\
		%	\hline
		%\end{tabular}
	%\end{center}
        %{\small
        \begin{tabular}{l|cccc|cccc|cccc}
		\hline
		block size & 5 & 10 & 25 & $\infty$ & 5 & 10 & 25 & $\infty$ & 5 & 10 & 25 & $\infty$ \\
		\hline
		\hline
		& \multicolumn{4}{c|}{RMSE of $\widehat\tht_{1:10}$} & \multicolumn{4}{c|}{RMSE of $\widehat\tht_{11:20}$} & \multicolumn{4}{c}{RMSE of $\widehat\rho$}\\
		\hline
		\multirow{2}*{$n=60$} &26&26&24&25&64&60&56&57&9&7&6&7\\
		 &25&25&23&24&61&57&52&53&7&6&5&5\\
		%$n=60$ &26&26&24&25&64&60&56&57&9.2&6.9&6.2&6.7\\
		%$n=60$ &25&25&23&24&61&57&52&53&6.8&6.0&5.1&5.2\\
		\hline
		\multirow{2}*{$n=200$} &16&14&13&14&39&28&26&26&6&5&4&3\\
		 &14&13&13&14&36&25&23&24&5&3&3&3\\
		%$n=200$ &16&14&&14&39&28&&26&6.4&4.5&&3.1\\
		%$n=200$ &14&13&&14&36&25&&24&5.2&3.2&&2.5\\
		\hline
		\hline
		& \multicolumn{4}{c|}{bias of $\widehat\tht_{1:10}$} & \multicolumn{4}{c|}{bias of $\widehat\tht_{11:20}$} & \multicolumn{4}{c}{bias of $\widehat\rho$}\\
		\hline
		\multirow{2}*{$n=60$} &$-8$&$-3$&$-2$&2&$-22$&$-10$&$-3$&5&6&4&2&$-1$\\
		 &$-5$&$-2$&$-2$&3&$-19$&$-9$&$-5$&3&4&2&1&$-1$\\
		%$n=60$ &8.4&3.1&1.9&2.1&22&10&3.2&4.9&5.7&4.1&2.2&0.6\\
		%$n=60$ &5.2&1.6&1.5&2.5&19&9.4&4.5&2.6&4.2&2.4&0.8&1.0\\
		\hline
		\multirow{2}*{$n=200$}&$-7$&$-4$&$-2$&1&$-26$&$-15$&$-7$&~~1&6&3&2&~~0\\
		&$-6$&$-2$&$-1$&1&$-24$&$-14$&$-7$&$-1$&4&2&1&$-1$\\
		%$n=200$ &7.3&3.5&&0.5&26&15&&0.6&5.8&3.3&&0.2\\
		%$n=200$ &5.6&2.4&&0.7&24&14&&0.7&4.4&2.1&&0.5\\
		\hline			
        \end{tabular}%}
        \end{center}
\end{table}

\subsection{Simulation 3: $\Sig$ has a block-diagonal structure, $d=15$} 

We simulate a data set from an extreme-value copula (\ref{eq-evcop}) with reflected Clayton linking copulas. We use $d=15$, and the vector of reflected Clayton linking copula parameters is $\tht = (2.5, 2.5, 2.5, 2.5, 2.5, 2.0, 2.0, 2.0, 2.0, 2.0, 1.5, 1.5, 1.5, 1.5, 1.5)^{\top}$. We select the correlation matrix $\Sig$ with a block-diagonal structure that has the off-diagonal elements  $\Sig_{j,k} = \rho_j\rho_k$ if $1 \leq j,k \leq 5$, $6 \leq j,k \leq 10$, $11 \leq j,k \leq 15$, and $\Sig_{j,k} = 0$ otherwise. One can check that $(\WW_1^{\top}, \WW_2^{\top}, \WW_3^{\top})^{\top} \sim N_{15}(\boldsymbol{0}, \Sig)$, where $\WW_g^{\top} = (W_{g,1}, \ldots, W_{g,5})$ and
$$
W_{g,i} = \rho_{g,i} Z_g + (1-\rho_{g,i}^2)^{1/2} Z_{g,i}, \quad g = 1, 2, 3, \ i = 1, \ldots, 5,
$$
where $Z_1, Z_2, Z_3, Z_{1,1}, \ldots, Z_{3,5} \sim_{\text{i.i.d.}} N(0,1)$. It implies that this correlation structure corresponds to the joint dependence of normal random variables from three independent groups, with the one-factor structure in each group. The respective copula (\ref{eq-evcop}) is an extreme-value limit of the bifactor copula model \citep{Krupskii.Joe2015b} with the reflected Clayton (normal) copulas linking the common (group-specific) factors, respectively, and the observed variables.

We use $\boldsymbol{\rho}_1 = (0.8, 0.6, 0.4, 0.2, 0.0)^{\top}$, $\boldsymbol{\rho}_2 = (0.4, 0.4, 0.4, 0.4, 0.4)^{\top}$ and $\boldsymbol{\rho}_3 = (0.0, 0.2, 0.4$, $0.6, 0.8)^{\top}$, where $\boldsymbol{\rho}_g = (\rho_{g,1}, \ldots, \rho_{g,5})^{\top}$. Similar to the previous two simulations, we use methods 1 and 2 to estimate the parameters $\tht$, $\boldsymbol{\rho}_1$, $\boldsymbol{\rho}_2$ and $\boldsymbol{\rho}_3$. Table \ref{tab3} shows the RMSEs of the estimates obtained by the two methods for 500 samples of size $m=60$ and $n=200$.

\begin{table}[h!]
	\caption{{\footnotesize RMSEs and biases (multiplied by 100) of the parameter estimates, $\widehat\tht$ and $\widehat{\boldsymbol{\rho}}$, obtained by method 1 (first line) and method 2 (second line). We used 500 samples of size $n=60$ and $n=200$ with reflected Clayton linking copulas and $\tht_{1:5}=2.5, \tht_{6:10}=2, \tht_{11:15}=1.5$, $\boldsymbol{\rho}_1 = (0.8, 0.6, 0.4, 0.2, 0.0)^{\top}$, $\boldsymbol{\rho}_2 = (0.4, 0.4, 0.4, 0.4, 0.4)^{\top}$ and $\boldsymbol{\rho}_3 = (0.0, 0.2, 0.4$, $0.6, 0.8)^{\top}$. }}
	\label{tab3}
	\begin{center}
		%{\footnotesize\begin{tabular}{lccccccccccccccc}
		%	\hline
		%	sample size & \multicolumn{15}{c}{RMSE of $\widehat\tht$ (top line) and $\widehat{\boldsymbol{\rho}}$ (bottom line)}\\
		%	\hline
		%	\multirow{2}{*}{$n=200$, method 1} &30&33&32&32&33&27&24&26&26&25&18&18&18&19&17\\
		%	&16&15&14&18&21&22&21&21&21&21&19&18&15&16&17\\
		%	\hline
		%	\multirow{2}{*}{$n=200$, method 2} &29&32&32&32&33&26&23&25&25&25&18&19&18&18&18\\
		%	&15&14&14&17&19&22&19&23&20&20&15&15&12&14&15\\
		%	\hline
		%	\multirow{2}{*}{$n=1000$, method 1} &15&14&14&14&13&10&11&11&10&11&7.7&7.5&8.1&7.7&7.9\\
		%	&7.6&6.3&6.2&7.3&9.2&8.6&8.8&8.5&8.3&8.4&7.5&6.4&6.0&6.9&8.5\\
		%	\hline
		%	\multirow{2}{*}{$n=1000$, method 2} &14&13&13&14&13&10&11&11&10&11&7.9&7.7&8.0&7.6&7.9\\
		%	&6.6&5.4&5.2&6.4&7.5&7.5&7.3&7.3&7.0&7.1&6.2&5.8&5.3&5.8&6.9\\
		%	\hline
		%\end{tabular}}
		\begin{tabular}{l|cccc|cccc|cccc}
			\hline
			block size & 5 & 10 & 25 & $\infty$ & 5 & 10 & 25 & $\infty$ & 5 & 10 & 25 & $\infty$ \\
			\hline
			\hline
			& \multicolumn{4}{c|}{RMSE of $\widehat\tht_{1:5}$} & \multicolumn{4}{c|}{RMSE of $\widehat\tht_{6:10}$} & \multicolumn{4}{c}{RMSE of $\widehat\tht_{11:15}$}\\
			\hline
			\multirow{2}*{$n=60$} &60&55&57&56&47&43&45&46&34&33&34&34\\
			&58&54&56&56&45&43&43&45&33&32&33&34\\
			\hline
			\multirow{2}*{$n=200$} &42&34&33&32&32&26&24&25&22&19&18&18\\
			&41&33&31&32&31&26&23&24&22&19&17&18\\
			\hline
			\hline
			& \multicolumn{4}{c|}{bias of $\widehat\tht_{1:5}$} & \multicolumn{4}{c|}{bias of $\widehat\tht_{6:10}$} & \multicolumn{4}{c}{bias of $\widehat\tht_{11:15}$}\\
			\hline
			\multirow{2}*{$n=60$} &$-32$&$-17$&$-7$&$-2$&$-19$&$-7$&$-1$&4&$-14$&$-7$&~~0&2\\
			&$-31$&$-19$&$-8$&$-5$&$-20$&$-10$&$-3$&1&$-13$&$-8$&$-1$&0\\
			\hline
			\multirow{2}*{$n=200$}&$-31$&$-15$&$-5$&~~0&$-23$&$-11$&$-3$&2&$-15$&$-8$&$-3$&1\\
			&$-30$&$-16$&$-7$&$-1$&$-22$&$-12$&$-5$&1&$-15$&$-8$&$-4$&0\\
			\hline	
			\hline
			& \multicolumn{4}{c|}{RMSE of $\widehat{\boldsymbol{\rho_1}}$} & \multicolumn{4}{c|}{RMSE of $\widehat{\boldsymbol{\rho_2}}$} & \multicolumn{4}{c}{RMSE of $\widehat{\boldsymbol{\rho_3}}$}\\
			\hline
			\multirow{2}*{$n=60$} &29&29&30&30&31&34&33&35&29&31&32&31\\
			&30&28&31&31&33&34&33&34&27&29&29&29\\
			\hline
			\multirow{2}*{$n=200$} &17&18&17&17&22&23&23&21&15&15&16&17\\
			&15&16&16&16&20&21&21&20&14&14&13&14\\
			\hline
			\hline
			& \multicolumn{4}{c|}{bias of $\widehat{\boldsymbol{\rho_1}}$} & \multicolumn{4}{c|}{bias of $\widehat{\boldsymbol{\rho_2}}$} & \multicolumn{4}{c}{bias of $\widehat{\boldsymbol{\rho_3}}$}\\
			\hline
			\multirow{2}*{$n=60$} &~~1&~~1&~~1&~~0&$-1$&$-3$&$-2$&$-4$&~~0&$-2$&$-3$&$-3$\\
			&$-1$&$-1$&$-2$&$-3$&$-3$&$-5$&$-4$&$-5$&$-1$&$-2$&$-3$&$-3$\\
			\hline
			\multirow{2}*{$n=200$}&~~0&~~0&~~0&~~0&0&$-2$&$-2$&$-1$&1&0&0&$-1$\\
			&$-1$&$-1$&$-1$&$-1$&0&$-2$&$-2$&$-2$&1&0&0&$-1$\\		
		\end{tabular}
	\end{center}
\end{table}

Similar to the previous simulations, there is a small bias when $n_B = 5$, and the results with $n_B = 25$ and $n_B = \infty$ are very close. The RMSEs of the estimates of $\tht$ and $\boldsymbol{\rho}$ are 5\% and 10--20\% higher for the fast method compared to the pairwise likelihood approach. Estimates of $\tht$ are less accurate if the dependence between the common factor and observed variables is stronger. Estimates of $\boldsymbol{\rho}$ are less accurate than estimates of $\tht$ but the accuracy improves with a larger sample size. 

\subsection{Simulation 4: $\Sig$ is a matrix of ones, $d=20$} 

We simulate a data set from an extreme-value copula such that $\Sig$ is a matrix of ones, and we use the Clayton linking copulas with continuous densities on $(0,1]^2$.  We use $d=20$, and the vector of Clayton linking copula parameters $\tht = (0.8,0.8,0.8,0.8,0.8,1.2,1.2,1.2,1.2,1.2,1.6$, $1.6,1.6,1.6,1.6,2,2,2,2,2)^{\top}$.
 
We use method 1 to estimate the copula parameters; see Appendix \ref{sec-appx2} for details. Parameters in this model are not identifiable and we therefore fix the first parameter at $0.8$. Table \ref{tab4} shows RMSEs of the copula parameter estimates obtained by method 1 for 500 samples of size $n=60$ and $n=200$.

%\begin{table}[h!]
%	\caption{{\footnotesize RMSEs (multiplied by 100) of the parameter estimates, $\widehat\tht_i$, $i=1,\ldots,10$ (first line) and $i=11,\ldots,20$ (second line), obtained by method 1. We used 500 samples of size $n=200$ and $n=1000$.  }}
%	\label{tab4}
%	\begin{center}
%		\begin{tabular}{lcccccccccc}
%			\hline
%			sample size & \multicolumn{10}{c}{RMSE of $\widehat\tht$}\\
%			\hline
%			\multirow{2}*{$n=200$} &0.0&0.0&0.0&0.0&0.0&5.6&5.6&5.6&5.6&5.6\\
%			&9.8&9.8&9.8&9.8&9.8&15&15&15&15&15\\
%			\hline
%			\multirow{2}*{$n=1000$} &0.0&0.0&0.0&0.0&0.0&2.3&2.3&2.3&2.3&2.3\\
%			&4.6&4.6&4.6&4.6&4.6&6.8&6.8&6.8&6.8&6.8\\
%			\hline
%		\end{tabular}
%	\end{center}
%\end{table}

\begin{table}[h!]
	\caption{{\footnotesize RMSEs and biases (multiplied by 100) of the parameter estimates, $\widehat\tht$, obtained by method 1. We used 500 samples of size $n=60$ and $n=200$ with Clayton linking copulas and $\tht_{1:5} = 0.8, \tht_{6:10}=1.2, \tht_{11:15} = 1.6, \tht_{16:20} = 2$. }}
	\label{tab4}
	\begin{center}
		{\small\begin{tabular}{l|cccc|cccc|cccc|cccc}
				\hline
				block size & 5 & 10 & 25 & $\infty$ & 5 & 10 & 25 & $\infty$ & 5 & 10 & 25 & $\infty$ & 5 & 10 & 25 & $\infty$ \\
				\hline
				\hline
				& \multicolumn{4}{c|}{RMSE of $\widehat\tht_{1:5}$} & \multicolumn{4}{c|}{RMSE of $\widehat\tht_{6:10}$} & \multicolumn{4}{c|}{RMSE of $\widehat\tht_{11:15}$} & \multicolumn{4}{c}{RMSE of $\widehat\tht_{16:20}$}\\
				\hline
				$n=60$ &0&0&0&0&11&11&11&10&21&21&23&21&32&33&35&32\\
				$n=200$&0&0&0&0&5&5&5&5&10&11&10&11&16&16&16&17\\
				\hline
				\hline
				& \multicolumn{4}{c|}{bias of $\widehat\tht_{1:5}$} & \multicolumn{4}{c|}{bias of $\widehat\tht_{6:10}$} & \multicolumn{4}{c|}{bias of $\widehat\tht_{11:15}$} & \multicolumn{4}{c}{bias of $\widehat\tht_{16:20}$}\\
				\hline
				$n=60$ &0&0&0&0&3&2&2&2&5&5&4&6&8&8&8&10\\
				$n=200$&0&0&0&0&1&1&1&1&1&2&1&1&2&3&2&2\\
				\hline			
		\end{tabular}}
	\end{center}
\end{table}

Once the first parameter is fixed, the remaining parameters can be identified. The estimates are less accurate for linking copulas with stronger dependence but RMSEs are quite small even for a small sample size $n = 60$. The results are very similar for different values of $n_B$ for this copula.

\subsection{Simulation 5: standard errors of the parameter estimators with known and estimated univariate marginals} 

We simulate two data sets from the same copulas as in Simulation 1 and consider two cases: 1) the univariate marginals are known so one can assume they follow a $U(0,1)$ distribution, and 2) the univariate marginals follow the Fr\'{e}chet distribution $F(z) = \exp\left\{-\left(\frac{z}{\sigma}\right)^{\alpha}\right\}$, $z > 0$ with $\alpha = 3$ and $\sigma = 1$. The parameters $\alpha$ and $\sigma$ are estimated using the maximum likelihood approach for each variable and the integral transform is used to transform the data to $U(0,1)$ marginals.

To obtain estimates of the exponent function required for method 1, we use the corrected estimator of the Pickands dependence function which yields more accurate estimates for distributions with known marginals \citep{Genest.Segers2009}.

We report the results for the block size $n_B = 25$ and $n_B = \infty$ and we use 500 samples of size $n=60$ and $n=200$. For the case of $n_B = 25$, we simulate data from the model (\ref{eq1_coppdf}) with the Pareto marginals with the cdf $F(z) = 1 - (\frac{z}{\sigma})^{-\alpha}$, $z > \sigma$ with $\alpha  = 3$ and $\sigma = 1$ before using block maxima. The exact univariate marginal distribution of the block maxima is therefore in the maximum domain of attraction of the Fr\'{e}chet distribution, so this distribution is used as an approximation. 
Table \ref{tab4b} show RMSEs of the copula parameter estimates obtained by methods 1 and 2 for the simulated data; we also include the results when the nonparamatric ranks are used to transform the data to uniform scores.

\begin{table}[h!]
	\caption{{\footnotesize RMSEs (multiplied by 100) of the parameter estimates, $\widehat\tht$ and $\widehat\rho$, obtained by method 1 / method 2 (first line: known marginals, second line: Fr\'{e}chet marginals with unknown parameters, third line: nonparametric ranks). We used 500 samples of size $n=60$ and $n=200$ with reflected Clayton copulas and $\tht_{1:3}=1, \tht_{4:7} = 2.5, \tht_{8:10}=1.5$ and $\Sig_{j,k} = \rho^{|j-k|}$ with $\rho=0.5$.  }}
	\label{tab4b}
	\begin{center}
		{\small\begin{tabular}{ll|cc|cc|cc|cc}
				\hline
				block size &  & 25 & $\infty$ & 25 & $\infty$ & 25 & $\infty$  & 25 & $\infty$ \\
				\hline
				\hline
				&marginals & \multicolumn{2}{c|}{RMSE of $\widehat\tht_{1:3}$} & \multicolumn{2}{c|}{RMSE of $\widehat\tht_{4:7}$} & \multicolumn{2}{c|}{RMSE of $\widehat\tht_{8:10}$} & \multicolumn{2}{c}{RMSE of $\widehat\rho$}\\
				\hline
				\multirow{3}*{$n=60$}& known &21/19&21/18&67/57&69/54&32/28&32/28&11/9&10/~8\\
				& Fr\'{e}chet & 25/23&22/21&71/63&71/63&34/34&35/34&10/9&10/~9\\
				& ranks & 24/23&24/24&76/74&78/74&36/34&38/36&11/9&11/10\\
				\hline
				\multirow{3}*{$n=200$} & known &11/10&11/10&31/27&31/26&15/14&16/15&~5/5&~5/~4\\
				&Fr\'{e}chet&12/11&12/12&36/31&32/30&18/17&17/17&~5/5&~5/~5\\
				&ranks&13/13&13/13&37/33&36/32&19/18&19/18&~6/5&~6/~5\\
				\hline			
		\end{tabular}}
	\end{center}
\end{table}

The RMSEs of the estimates of $\tht$ and $\rho$ are smallest both for methods 1 and 2 if univariate marginals are known but the loss of efficiency is not very large with unknown marginals. In particular, the RMSEs are 10-15\% higher for method 1 and about 20\% higher for method 2 when nonparametric ranks are used, and 5-10\% higher with estimated Fr\'{e}chet marginals. 

Method 2 yields more accurate estimates, with about 10\% smaller RMSEs as compared to method 1. Both methods have comparable RMSEs with $n_B = 25$ and $n_B = \infty$ which indicates that the block size $n_B = 25$ is large enough to assume the extreme-value limit of the model (\ref{eq1_coppdf}) is a good approximation to the block maxima data generated from this model. 
We obtained similar results for other models including those we used in simulations 2, 3 and 4, and with different parameter values. 

\section{Empirical Studies}
\label{sec_empstudy}

In this section, we apply the conditional normal extreme-value copulas to analyze two data sets. The first data set consists of the monthly maxima of daily average wind speed data, and the second data set is made of the monthly minima of daily stock log-returns. In applications, modeling the wind data can help wind turbine operators to predict downtimes due to very strong winds, and a model for the stock log-returns can be used by an investor to evaluate the performance of their portfolio under negative scenarios when extreme comovements in stock markets are observed within a short period of time.
\subsection{Wind data}

We consider monthly maxima of the average daily wind speed at a height of 10 meters above the ground recorded by 12 weather stations (Acme, Burneyville, Byars, Centrahoma, Chickasha, Ketchum Ranch, Lane, Ringling, Shawnee, Sulphur, Washington, and Waurika) in Oklahoma state, United States. The data are available at the \texttt{mesonet.org} website.

The stations are located within a small geographical region, the maximum distance between two stations being 194 km, and at an elevation between 181 and 430 m. We therefore assume spatial stationarity for these data. The average wind speed is higher in winter and spring due to thunderstorms, which produce gusty winds, and it is lower in summer when the weather is more settled. We therefore include three summer months (June, July and August) from 2000 to 2019, 60 months in total. 

We use  model (\ref{eq-evcop}) with the reflected Clayton copulas with the same parameter $\theta$ and spatial correlation matrix $\Sig$ with $\Sig_{j,k} = (1-\eta)\exp(-\gamma r_{j,k}^{\alpha})$, where the parameter $0 < \eta < 1$ is used to model the nugget effect, $\gamma > 0$ and $0 < \alpha \leq 2$ are parameters of the powered-exponential covariance function, and $r_{j,k}$ is the distance (in km) between the $j$th and $k$th stations.   

\begin{figure}[h!]
	\begin{center}\vspace{-1.4cm}
		\includegraphics[width=6.0in,height=3.5in]{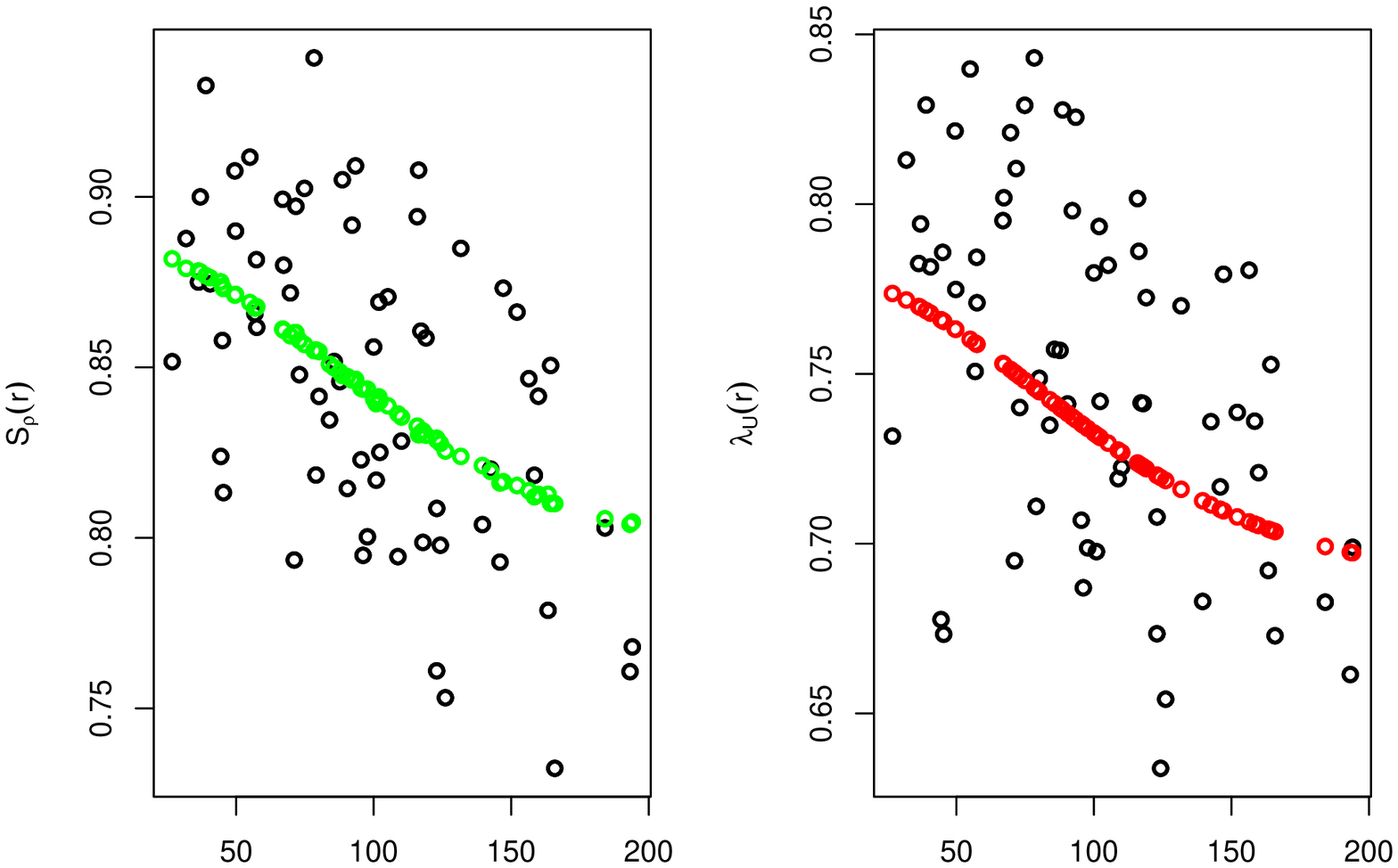}\vspace{-1cm}
		\includegraphics[width=6.0in,height=3.5in]{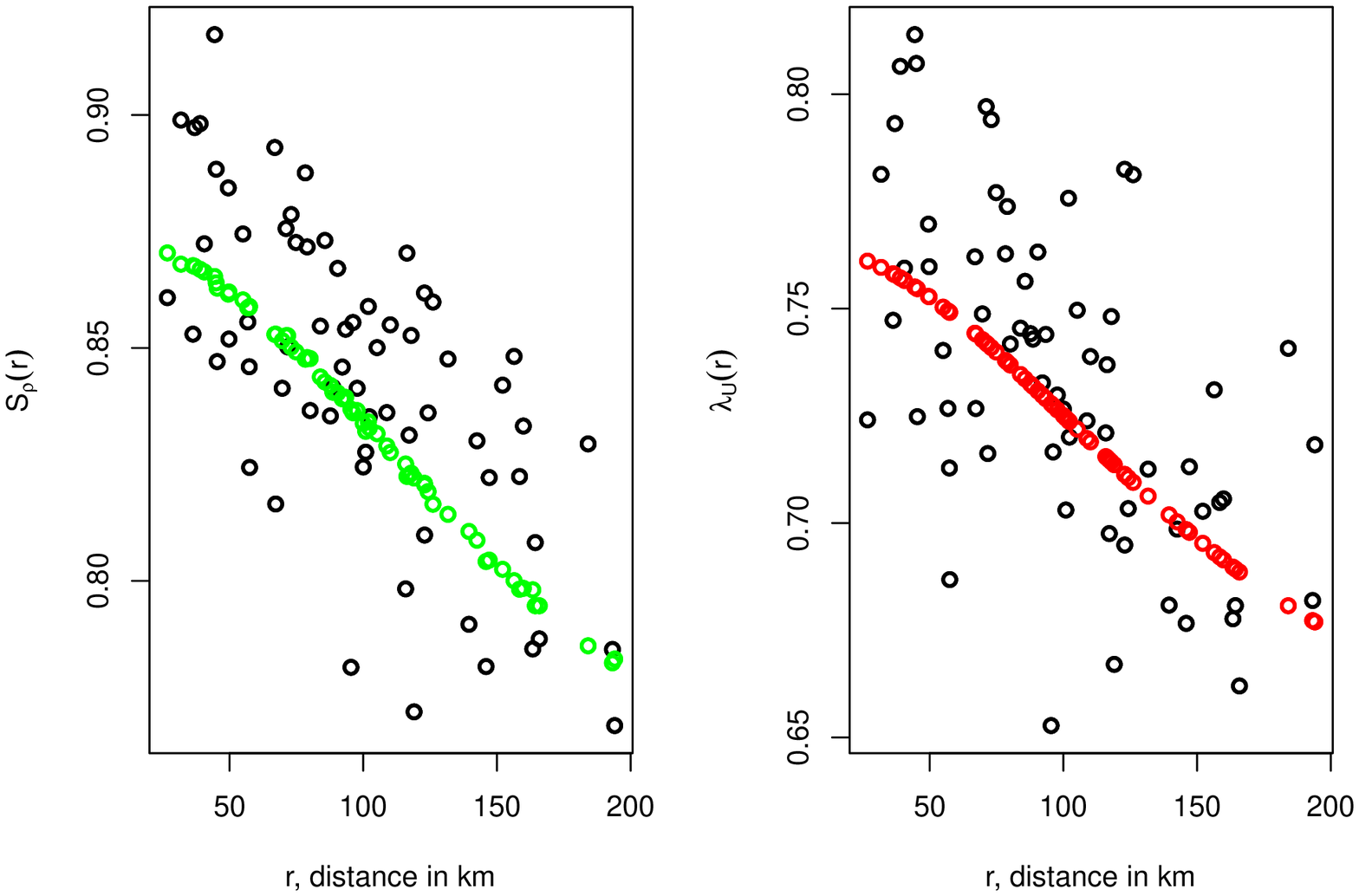}\vspace{-0.6cm}
		\caption{{\footnotesize Empirical and the estimated correlations, $S_{\rho}(r)$ (black and green points), and empirical and estimated upper tail dependence coefficients, $\lambda_U(r)$ (black and red points), for different pairs of stations of the Oklahoma wind data (first row) and pairs of variables from the simulated data set (second row)}}
		\label{fig_spat}
	\end{center}
\end{figure}

We transform the measurements taken at each station into uniform data using nonparametric ranks, and we estimate the model parameters using the pairwise likelihood approach. We get the following estimates: 
$$\widehat\theta = 2.82, \quad \widehat\eta = 0.49, \quad \widehat\gamma = 6350, \quad \widehat\alpha = 2.$$ Next, we compute the empirical and model-based estimates of Spearman's correlation $S_{\rho}$ and the upper tail dependence coefficient $\lambda_U$ for different pairs of stations. For the model-based estimates, the formula for $V_{j,k}(w_j,w_k)$ can be used to compute $\lambda_U = 2 - V_{j,k}(1,1)$ for the $(j,k)$ pair of variables; see Appendix \ref{sec-appx2}. For the empirical estimates, we use formula (\ref{eq-lambda}). There is no explicit formula for $S_{\rho}$, so we simulate a sample of size 10000 from the estimated model to compute the model-based estimates of correlations. The empirical and model-based estimates of $S_{\rho} = S_{\rho}(r)$ and $\lambda_U = \lambda_U(r)$ are plotted against distance $r$ in Figure \ref{fig_spat} (first row).

%As shown in Figure \ref{fig_spat}, 
Although the variability of empirical estimates is high due to a small sample size, the empirical and estimated values are in good agreement. %conditional normal extreme-value copula fits the data well. 
The relationship between $\lambda_U(r)$ and $r$ is approximately linear for a small $r$, which suggests that $(\eta + Cr^{\alpha})^{1/2} = \eta^{1/2}+O(r)$ or $\alpha = 2$ under the assumptions of Proposition \ref{prop6}.  The estimate $\widehat\alpha=2$ is therefore in agreement with the result of this proposition.

\begin{figure}[h!]
	\begin{center}\vspace{-1.4cm}
		\includegraphics[width=6.0in,height=2.5in]{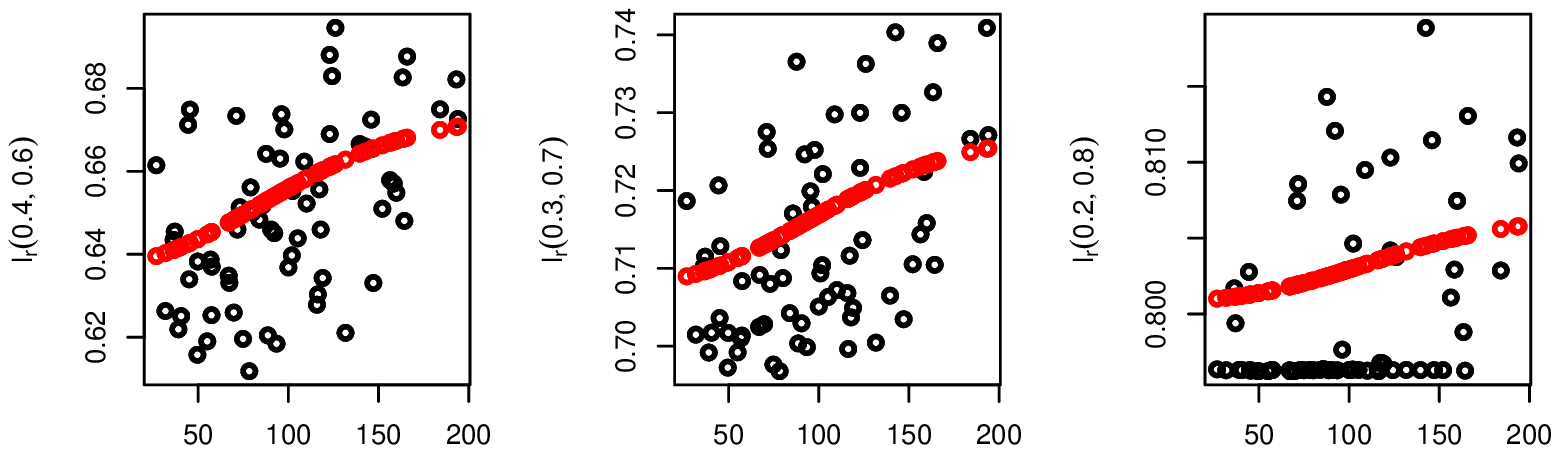}\vspace{-1.5cm}
		\includegraphics[width=6.0in,height=2.5in]{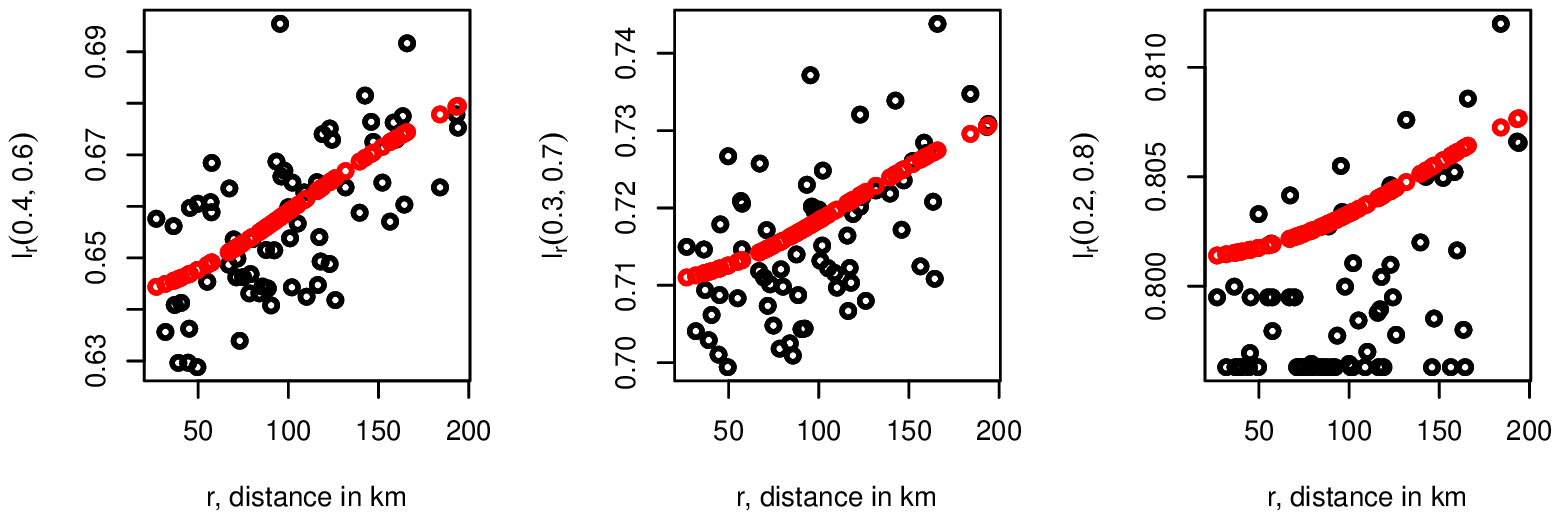}\vspace{-0.6cm}
		\caption{{\footnotesize Empirical and the estimated tail dependence functions, $\ell_r(w, 1-w)$ (black and red points), for different pairs of stations of the Oklahoma wind data (first row) and pairs of variables from the simulated data set (second row)}}
		\label{fig_spat2}
	\end{center}
\end{figure}

We also calculate the empirical estimates of the tail dependence function $\ell_r(w, 1-w)$ for pairs of stations at distance $r$ using the nonparametric estimator proposed by \cite{Genest.Segers2009}, and we compare them against the model-based estimates. Figure \ref{fig_spat2} (first row) shows the results for different pairs of stations and $w = 0.2, 0.3, 0.4$; we also obtained similar results for $w = 0.6, 0.7, 0.8$.

Again, we can see that the empirical and estimated values are close pointing to a good fit of the proposed model to the wind data. We obtain similar results for different values of~$w$.

Finally, we check if the selected model is robust to linking copulas misspecification. We generate a sample of size $n=60$ from an extreme-value copula (\ref{eq-evcop}) with the Gumbel linking copulas. To replicate the dependence structure of the wind data, we use the following parameters
$$
\theta = 3.3, \quad \eta = 0.49, \quad \gamma=6.35\cdot 10^{-5}, \quad \alpha =2.
$$  
 
With these parameters, the overall strength of dependence for different pairs of variables (as measured by $S_{\rho}$) is close to the empirical correlations for the original data set. 
We then use model (\ref{eq-evcop}) with the misspecified reflected Clayton copulas and spatial correlation matrix $\Sig$ as before. For the estimated model, we compute the empirical and estimated values of $S_{\rho}(r), \lambda_U(r)$ and $\ell_r(w, 1-w)$ for $w=0.2, 0.3, 0.4$; Figures %\ref{fig_spat_ms} and \ref{fig_spat2_ms} show the results. 
\ref{fig_spat} and \ref{fig_spat2} (second rows) show the results.
%\begin{figure}[h!]
%	\begin{center}\vspace{-1.4cm}
%		\includegraphics[width=6in,height=3.5in]{fig_emp_study1_missp1c.eps}\vspace{-0.6cm}
%		\caption{{\footnotesize Empirical and the estimated correlations, $S_{\rho}(r)$ (black and green points), and empirical and estimated upper tail dependence coefficients, $\lambda_U(r)$ (black and red points), for different pairs of variables from the simulated data set}}
%		\label{fig_spat_ms}
%	\end{center}
%\end{figure}
%
%\begin{figure}[h!]
%	\begin{center}\vspace{-1.4cm}
%		\includegraphics[width=6.0in,height=2.5in]{fig_emp_study1_missp2.eps}\vspace{-0.6cm}
%		\caption{{\footnotesize Empirical and the estimated tail dependence functions, $\ell_r(w, 1-w)$ (black and red points), for different pairs of variables from the simulated data set}}
%		\label{fig_spat2_ms}
%	\end{center}
%\end{figure}

The empirical and model-based estimates are quite close, and this indicates that the misspecified model has a reasonably good fit to the simulated data.
 
\subsection{Stock return data}

We use monthly minima of daily stock log-returns from the S\&P 500 index. We include stocks from three sectors: seven stocks from the industrial machinery sector with tickers \texttt{CMI, DOV, GWW, ITW, PH, SNA, SWK}; ten stocks from the electric utilities sector with tickers \texttt{AEP, ED, D, DUK, EIX, ETR, FE, PPL, PEG, SO}; and seven stocks from the regional banks sector with tickers \texttt{FITB, HBAN, KEY, MTB, PNC, RF, TFC}. The study period is 2000--2019, 240 months in total. %As we focus on the models with static dependence in this paper, we assume stationarity of the stock return data. We therefore exclude 2007-2009 when the US subprime mortgage crisis severely affected the US market and resulted in a much stronger dependence among extreme returns. 

We use the model (\ref{eq-evcop}) with the reflected Clayton copulas with a vector of parameters $\tht$ and the correlation matrix $\Sig$ with a block-diagonal structure with a vector of correlation parameters $\boldsymbol{\rho}$ and the off-diagonal elements $\Sig_{j,k} = \rho_j\rho_k$ if a pair $(j,k)$ is from the same sector, and $\Sig_{j,k} = 0$ otherwise. This corresponds to a model with the bifactor structure as in simulation 3 in Section \ref{sec_simul}. Bifactor models can handle data with factor structures, and can be suitable for modeling financial data with stock returns affected by global macroeconomic and sector-specific factors.

We transform the log-returns minima to uniform data using nonparametric ranks and fit the proposed model to the reflected data using the pairwise likelihood approach. The parameter estimates for the three groups are

{\small \begin{eqnarray*}
\widehat\tht_{1:7} &=& (1.46, 1.86, 0.78, 1.41, 1.27, 1.18, 0.97)^{\top},\\
\widehat{\boldsymbol{\rho}}_{1:7} &=& (0.06, -0.24, 0.19, 0.78, 0.39, 0.29, 0.63)^{\top},\\
\widehat\tht_{8:17} &=& (0.72, 0.65, 0.71, 0.53, 0.55, 0.74, 0.62,
0.77, 0.79, 0.63)^{\top},\\
\widehat{\boldsymbol{\rho}}_{8:17} &=& (0.92, 0.87, 0.89, 0.96, 0.86, 0.84, 0.69, 0.86, 0.83, 0.89)^{\top},\\
\widehat\tht_{18:24} &=& (1.08, 1.11, 1.21, 1.15, 1.49, 0.95, 1.21)^{\top},\\
\widehat{\boldsymbol{\rho}}_{18:24} &=& (0.87, 0.90, 0.90, 0.78, 0.73, 0.94, 0.86)^{\top}.\\
\end{eqnarray*}}
\vspace{-1.2cm}

The results indicate that the dependence between the observed variables and the common factor is stronger in the first and third sectors, while the group-specific factors are strongly correlated with the variables from the second and third groups. This implies that sectors 1 and 3 are strongly dependent on each other and that the residual dependence of these data is very strong, so a model with only one common factor is not suitable for the monthly minima of log-returns.

\begin{figure}[h!]
	\begin{center}%\vspace{-1.4cm}
		\includegraphics[width=1.8in,height=1.9in]{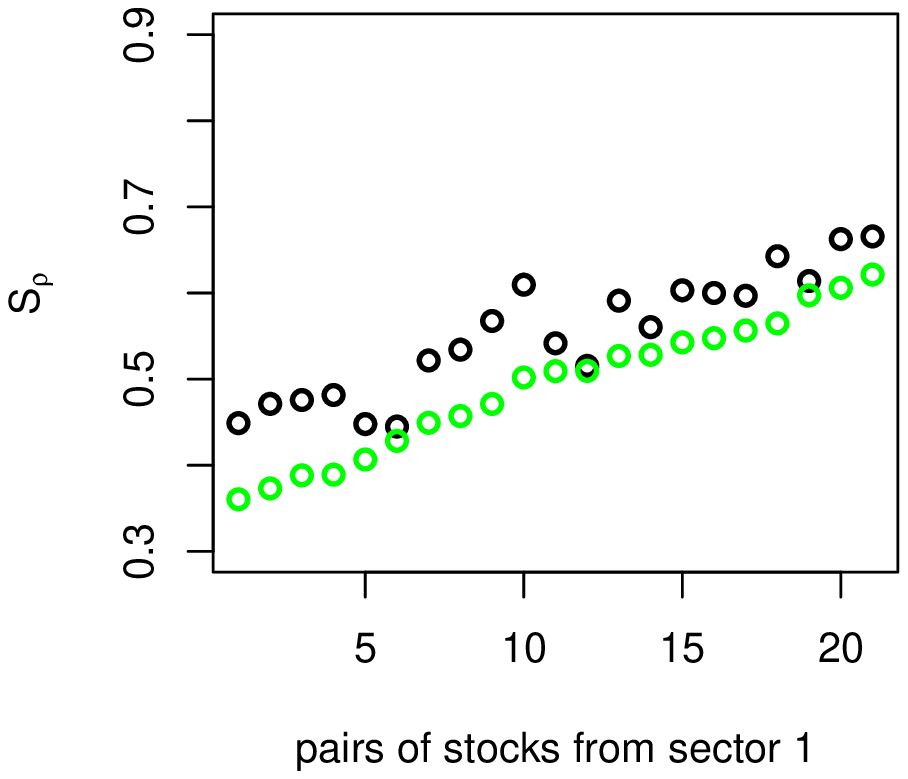}%\vspace{-0.6cm}
		\includegraphics[width=1.8in,height=1.9in]{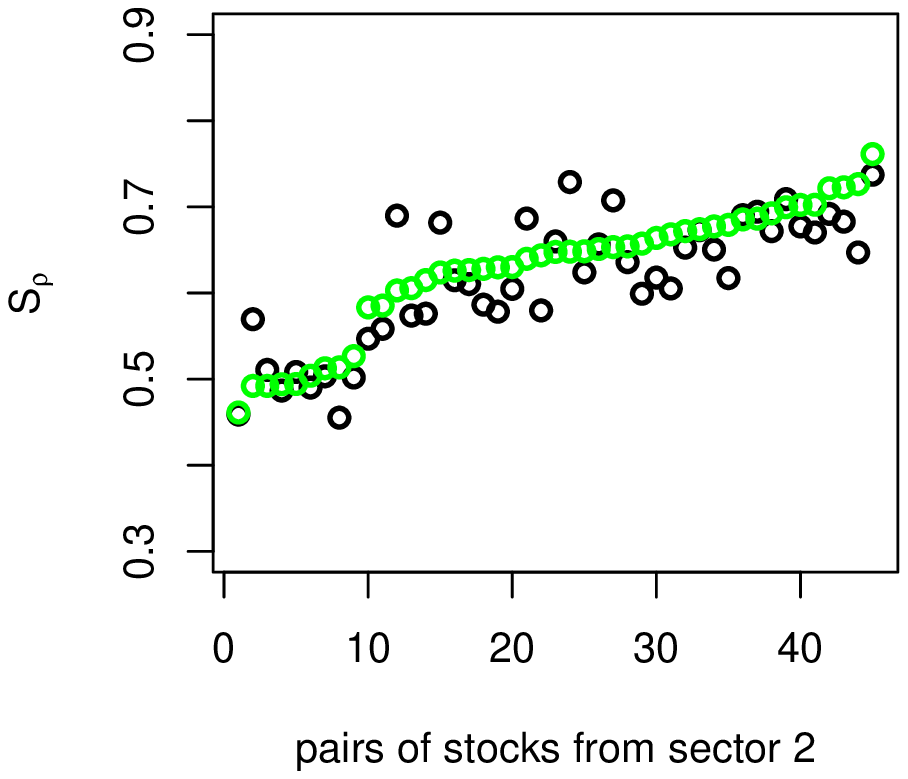}%\vspace{-0.6cm}
		\includegraphics[width=1.8in,height=1.9in]{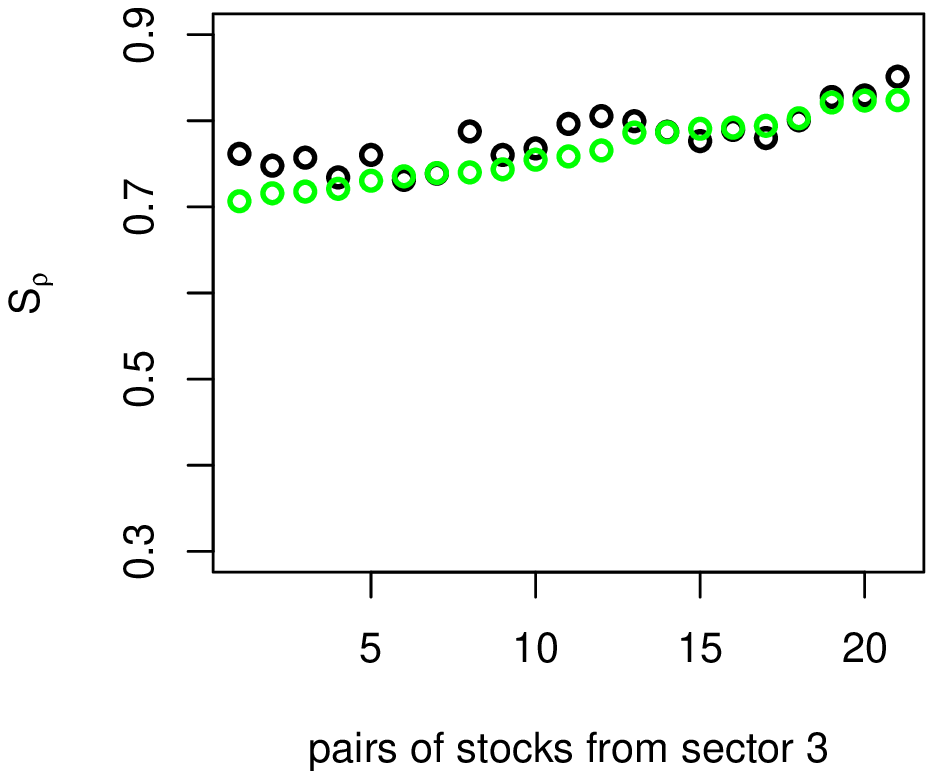}\\
		\vspace{-0.7cm}
		\includegraphics[width=1.8in,height=1.9in]{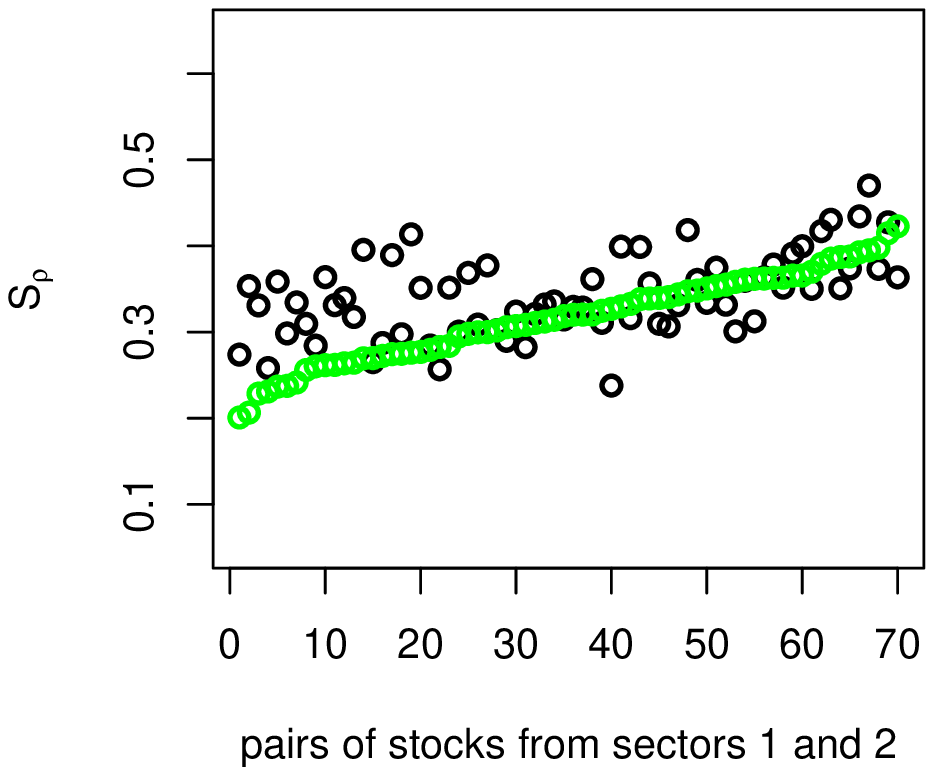}%\vspace{-0.6cm}
		\includegraphics[width=1.8in,height=1.9in]{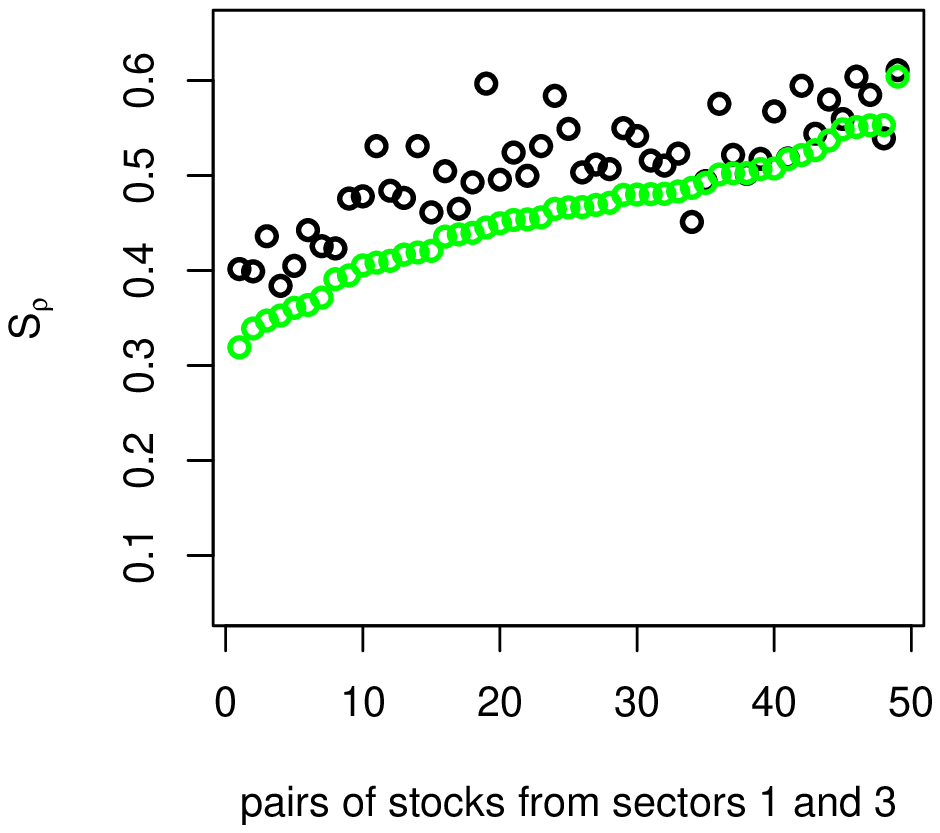}%\vspace{-0.6cm}
		\includegraphics[width=1.8in,height=1.9in]{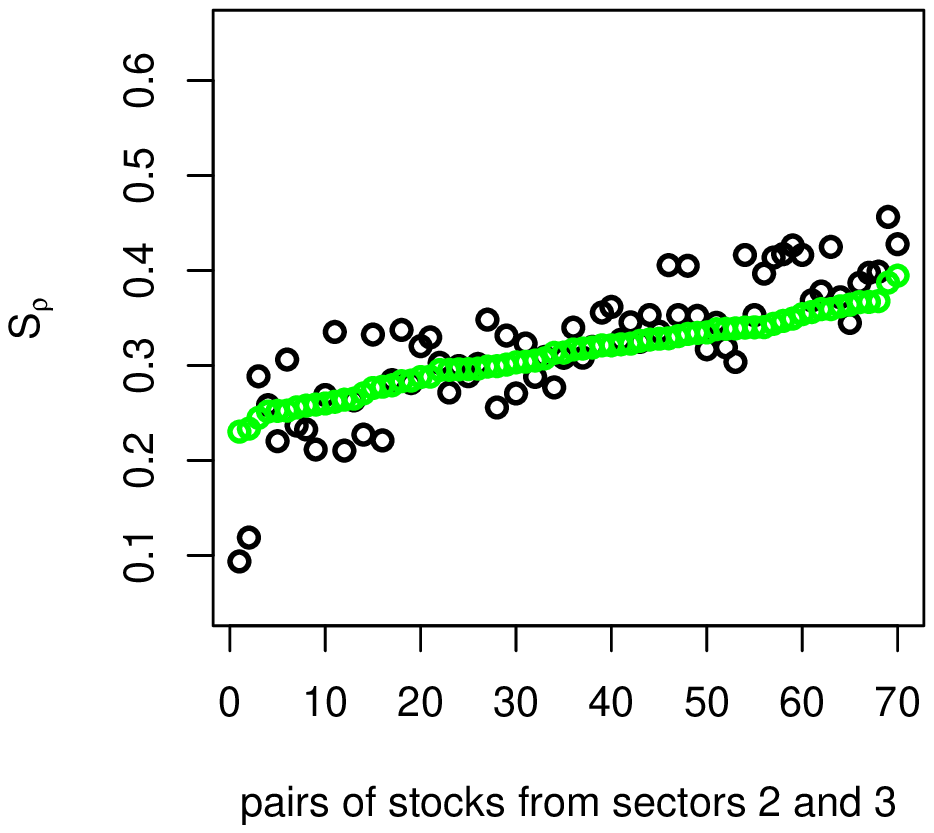}\\%\vspace{-0.6cm}
		\vspace{-0.7cm}
		\includegraphics[width=1.8in,height=1.9in]{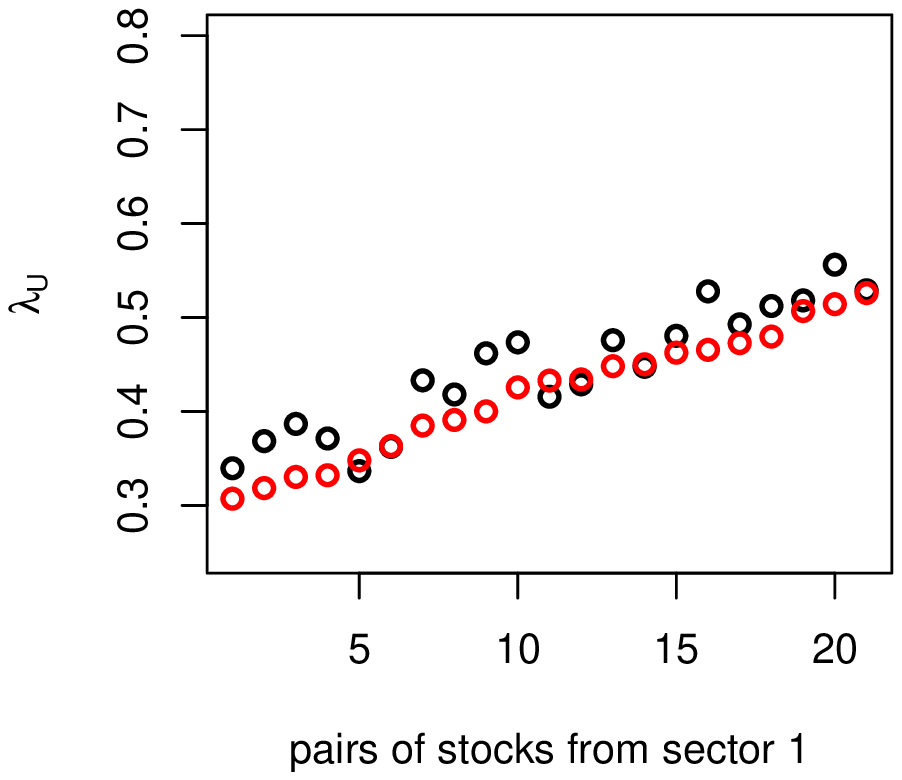}%\vspace{-0.6cm}
		\includegraphics[width=1.8in,height=1.9in]{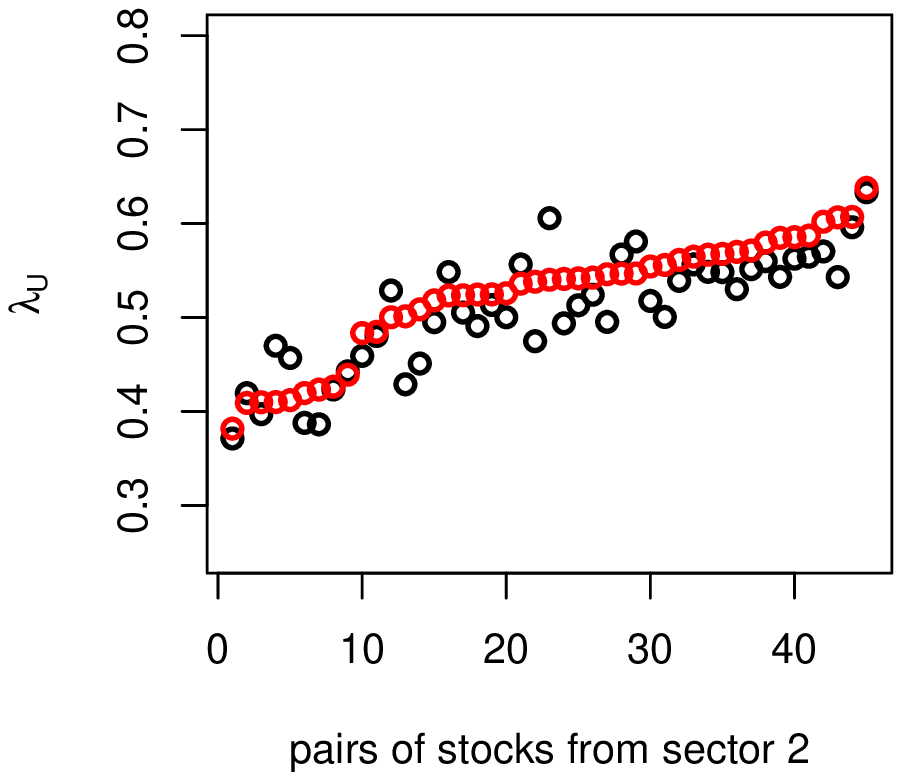}%\vspace{-0.6cm}
		\includegraphics[width=1.8in,height=1.9in]{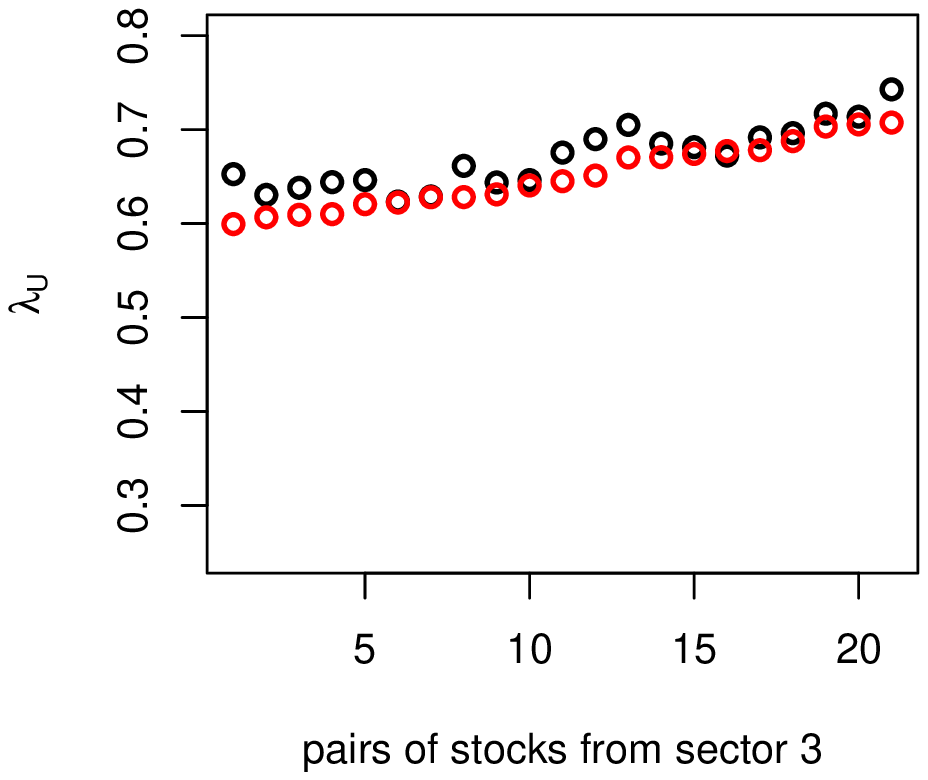}\\
		\vspace{-0.7cm}
		\includegraphics[width=1.8in,height=1.9in]{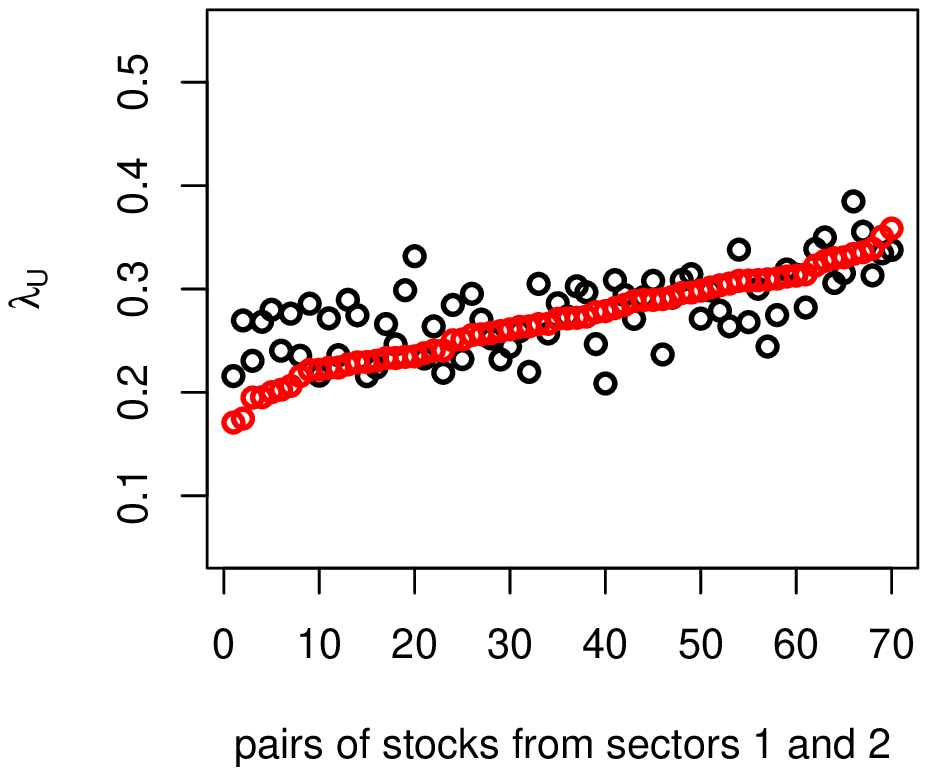}%\vspace{-0.6cm}
		\includegraphics[width=1.8in,height=1.9in]{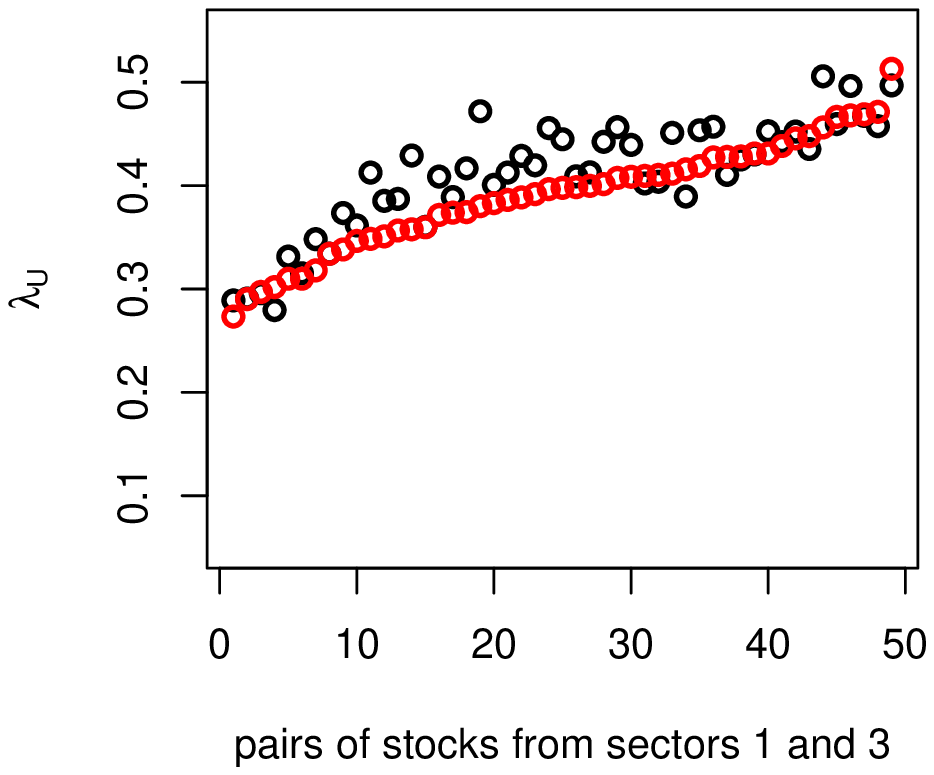}%\vspace{-0.6cm}
		\includegraphics[width=1.8in,height=1.9in]{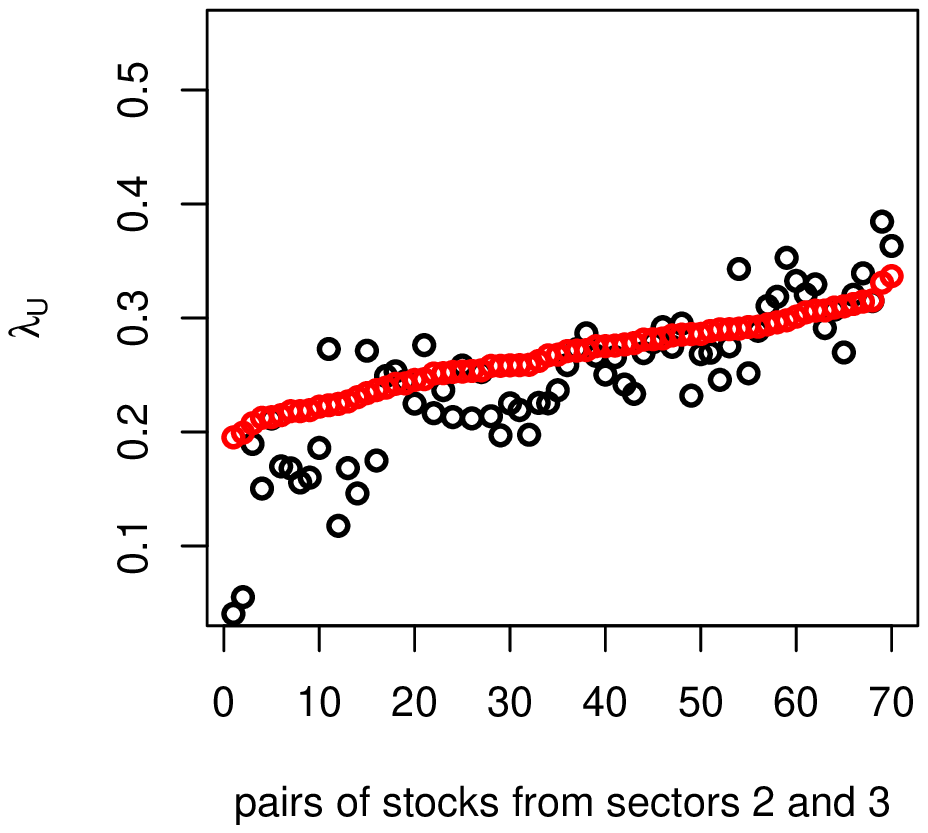}%\vspace{-0.6cm}
		\caption{{\footnotesize Empirical and estimated correlations, $S_{\rho}$ (first two rows, black and green points), and upper tail dependence coefficients, $\lambda_U$ (last two rows, black and red points), for different pairs of stock returns  from sector 1 (top left), sector 2 (top middle), sector 3 (top right), and sectors 1, 2 (bottom left), sectors 1, 3 (bottom middle), sectors 2, 3 (bottom right) }}
		\label{fig_1fct}
	\end{center}
\end{figure}

Similar to the wind data, we compute the empirical and model-based estimates of Spearman's correlation $S_{\rho}$ and the upper tail dependence coefficient $\lambda_U$ for different pairs of variables from the same sectors and from different sectors. To make comparisons easier, we sort the model-based estimates in an increasing order; Figure \ref{fig_1fct} shows the results.

As shown in Figure \ref{fig_1fct}, dependence within the sector, as measured by $S_{\rho}$ and $\lambda_U$, is stronger than dependence between the sectors, as expected. The dependence between sectors 1 and 3 is slightly weaker than the dependence within sector 1, and the dependence between sectors 1 and 2 and between sectors 2 and 3 is significantly weaker than the dependence within respective sectors, pointing to a weaker dependence between sectors 1 and 2, and sectors 2 and 3. 

\begin{figure}[h!]
	\begin{center}%\vspace{-1.4cm}
		\includegraphics[width=1.8in,height=1.9in]{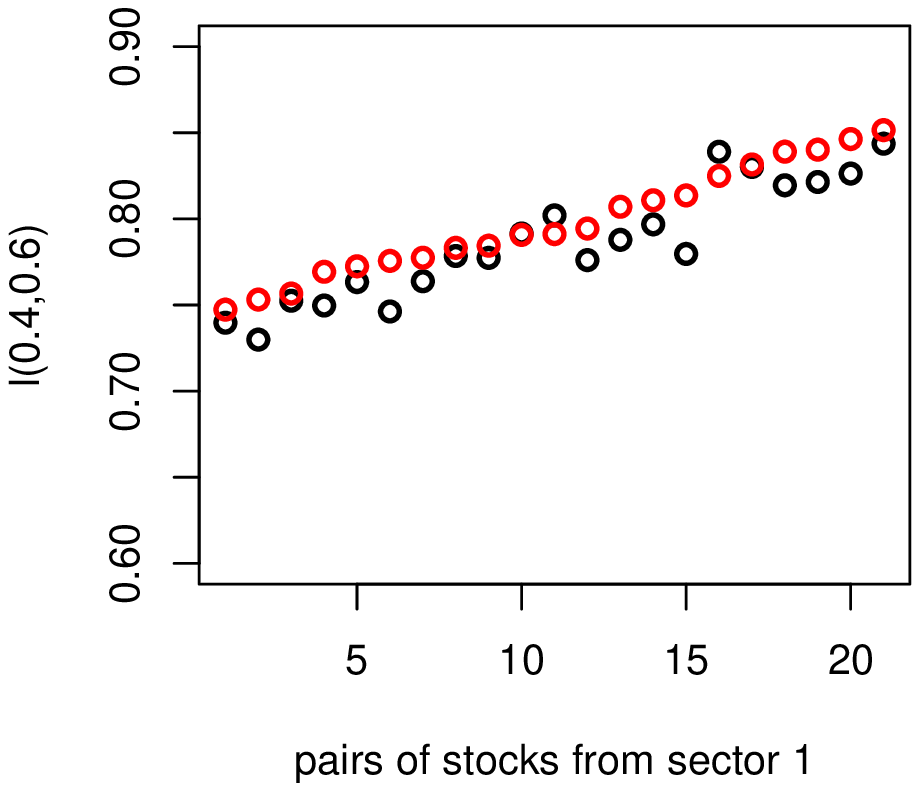}%\vspace{-0.6cm}
		\includegraphics[width=1.8in,height=1.9in]{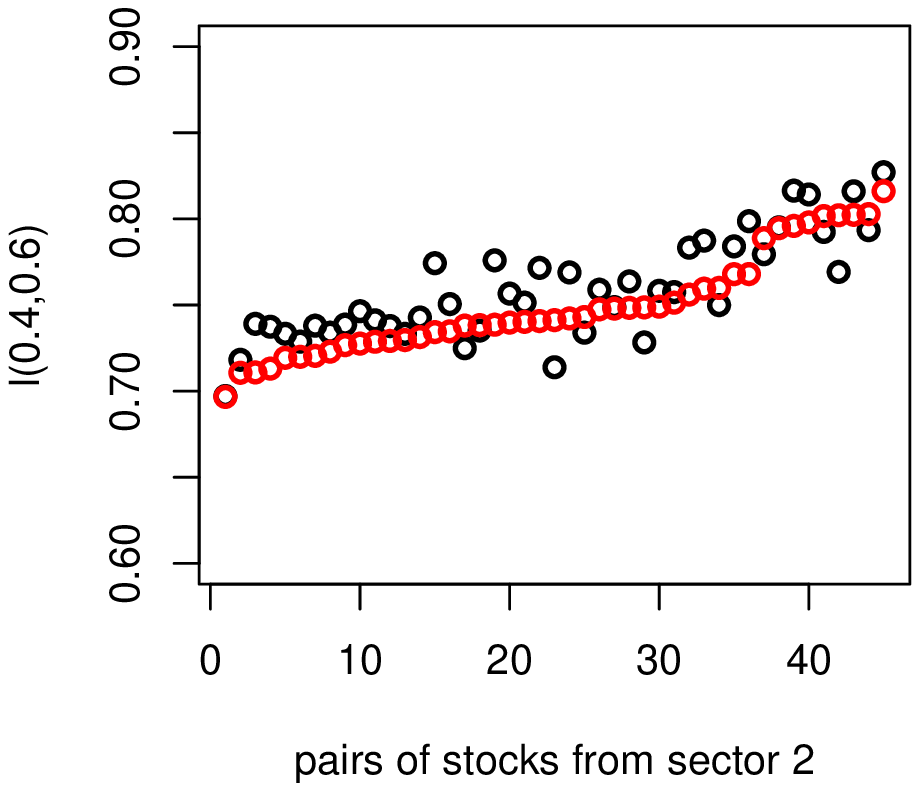}%\vspace{-0.6cm}
		\includegraphics[width=1.8in,height=1.9in]{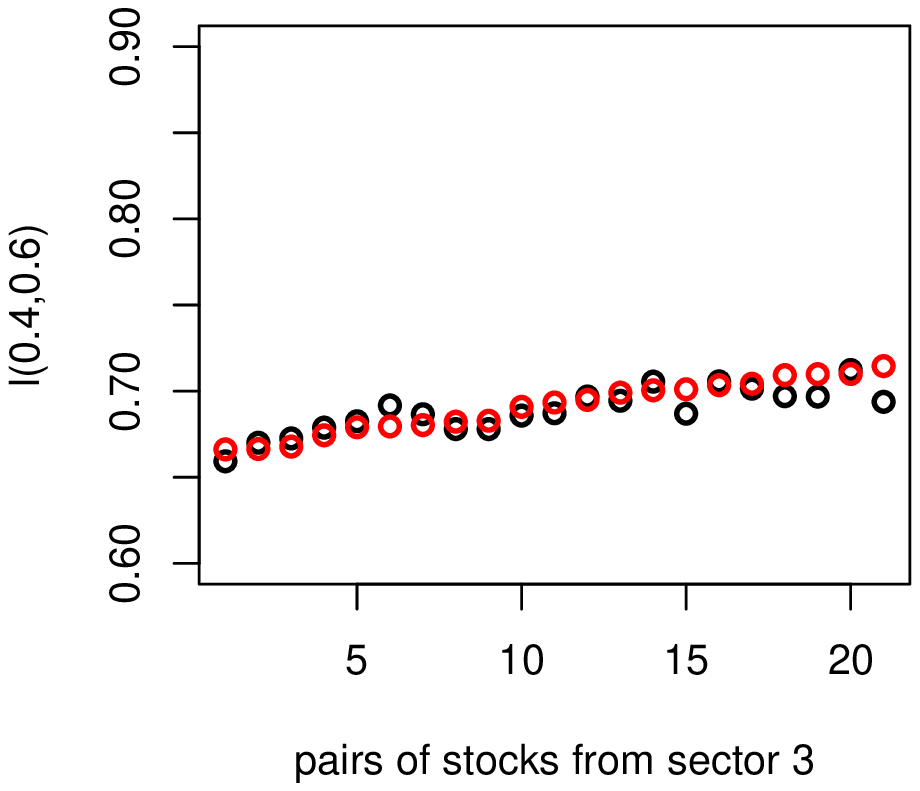}\\
		\vspace{-0.7cm}
		\includegraphics[width=1.8in,height=1.9in]{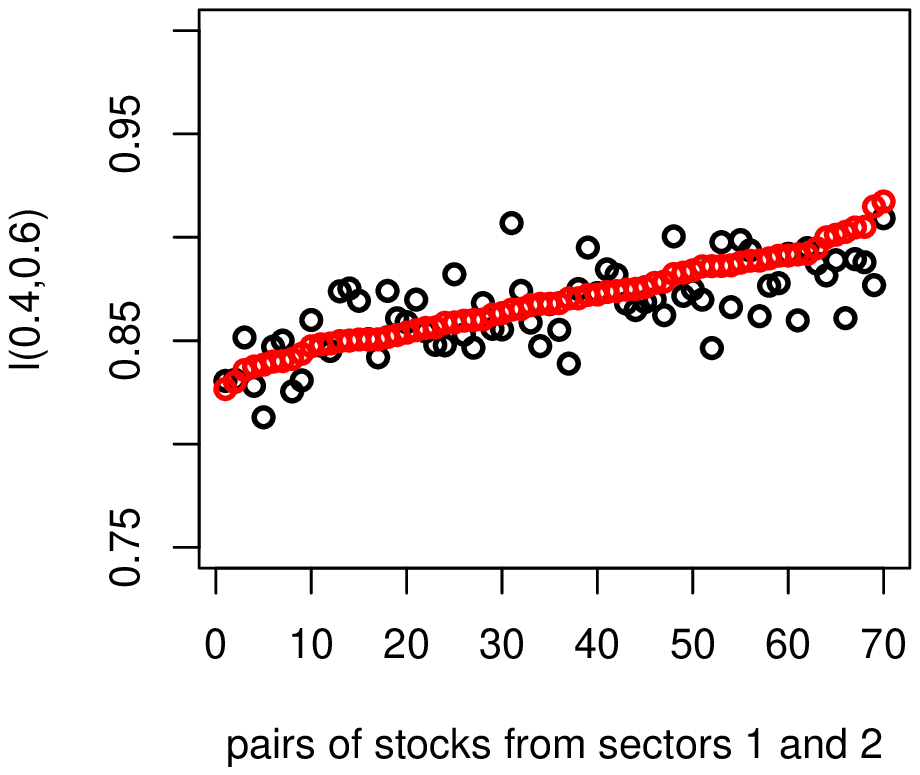}%\vspace{-0.6cm}
		\includegraphics[width=1.8in,height=1.9in]{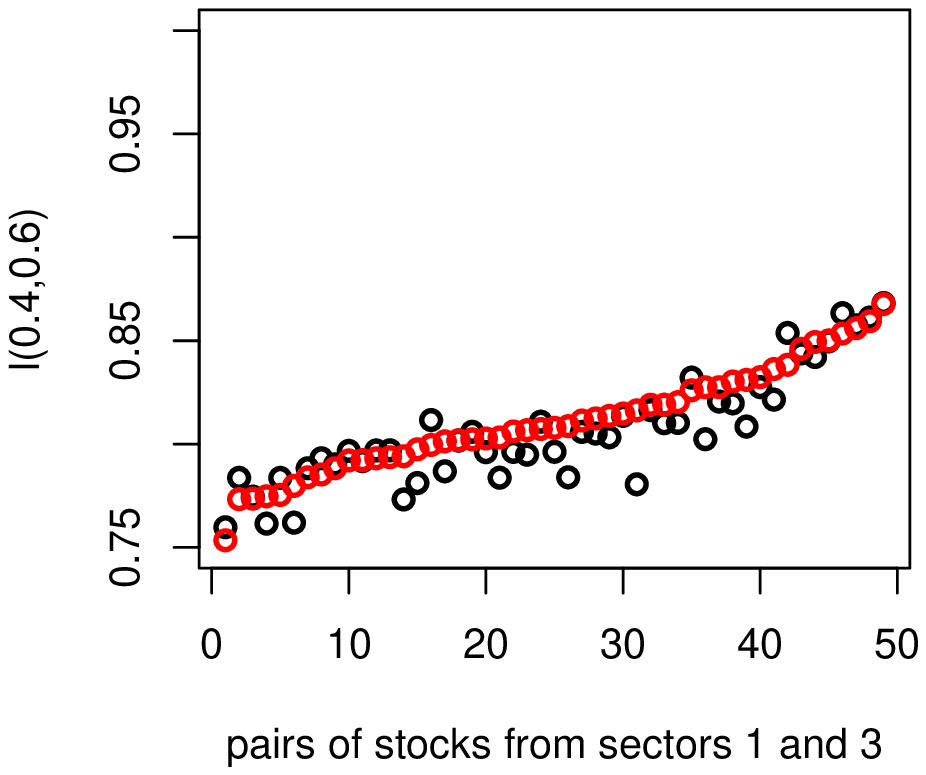}%\vspace{-0.6cm}
		\includegraphics[width=1.8in,height=1.9in]{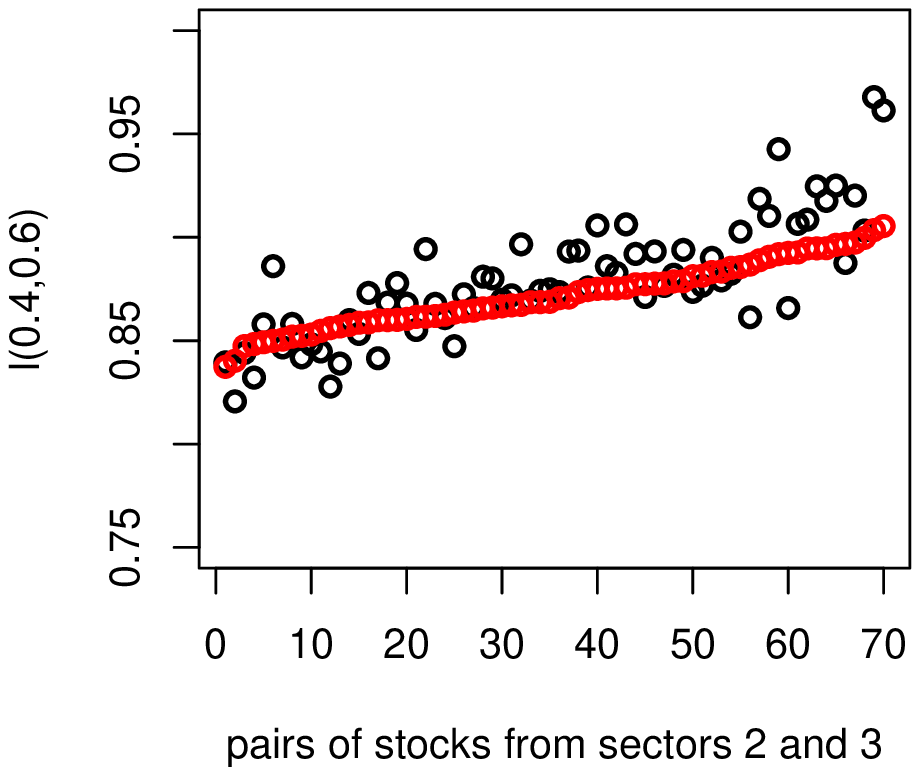}\\%\vspace{-0.6cm}
		\vspace{-0.7cm}
		\includegraphics[width=1.8in,height=1.9in]{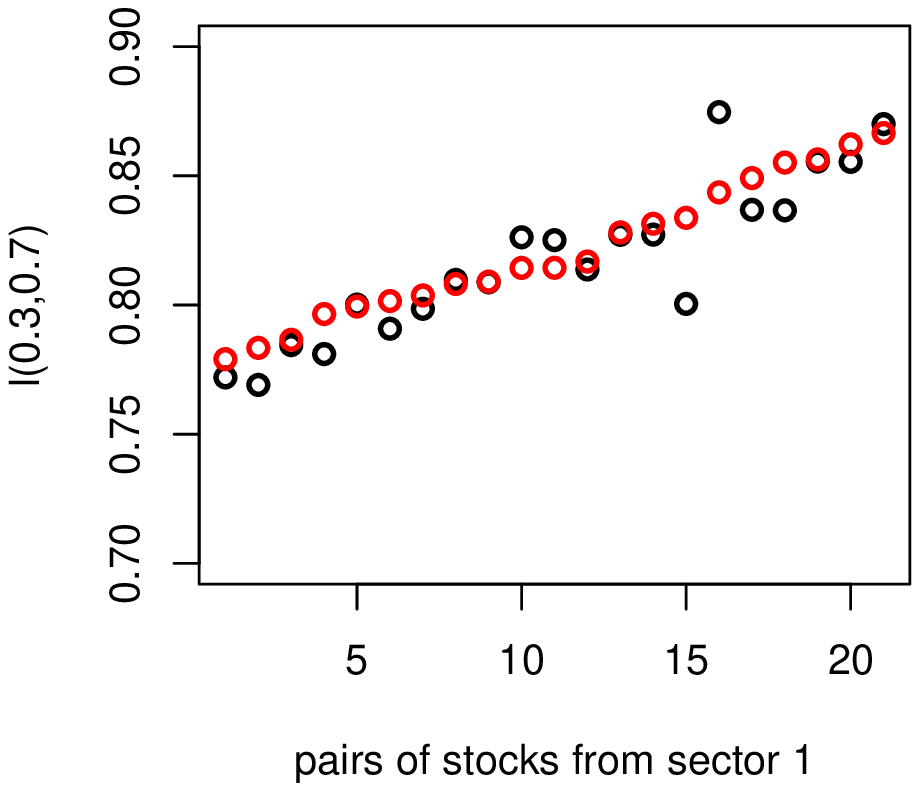}%\vspace{-0.6cm}
		\includegraphics[width=1.8in,height=1.9in]{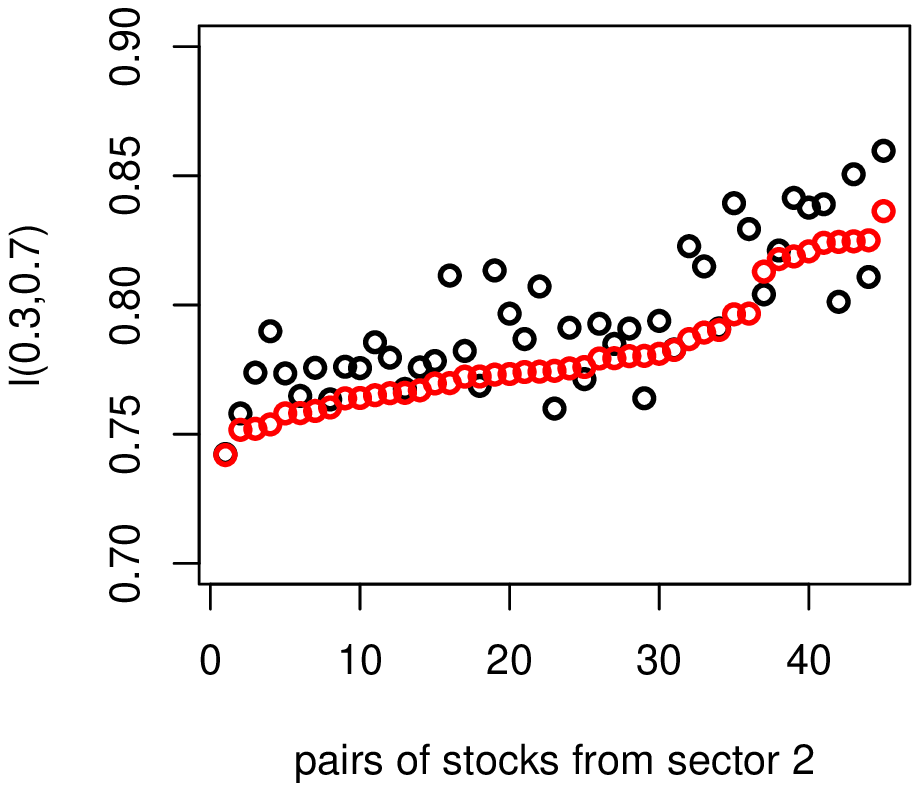}%\vspace{-0.6cm}
		\includegraphics[width=1.8in,height=1.9in]{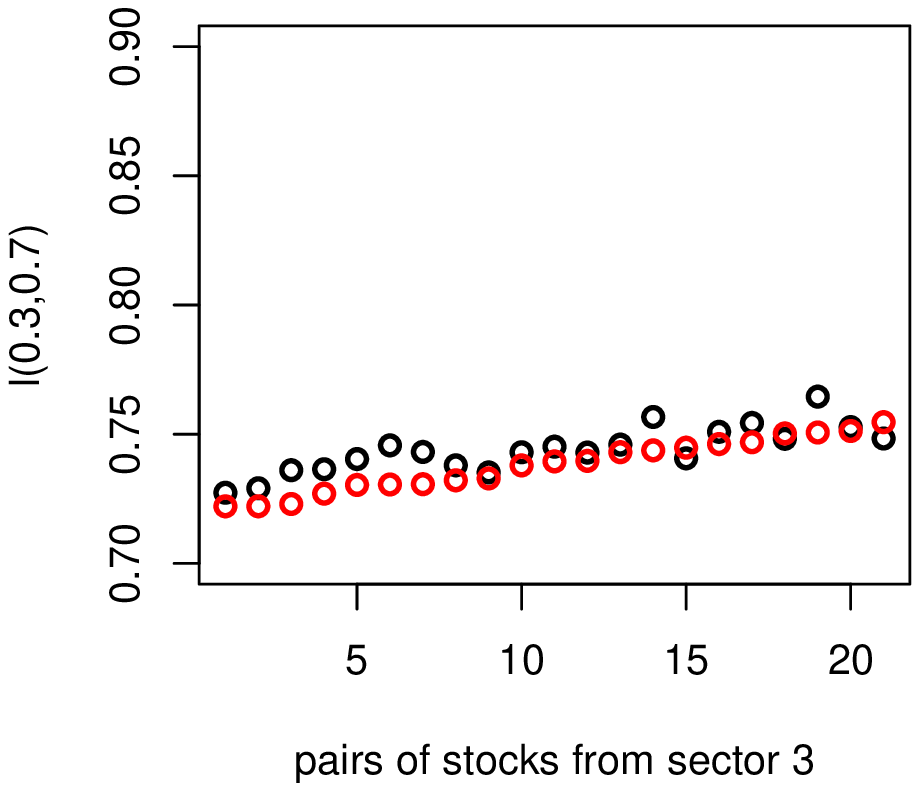}\\
		\vspace{-0.7cm}
		\includegraphics[width=1.8in,height=1.9in]{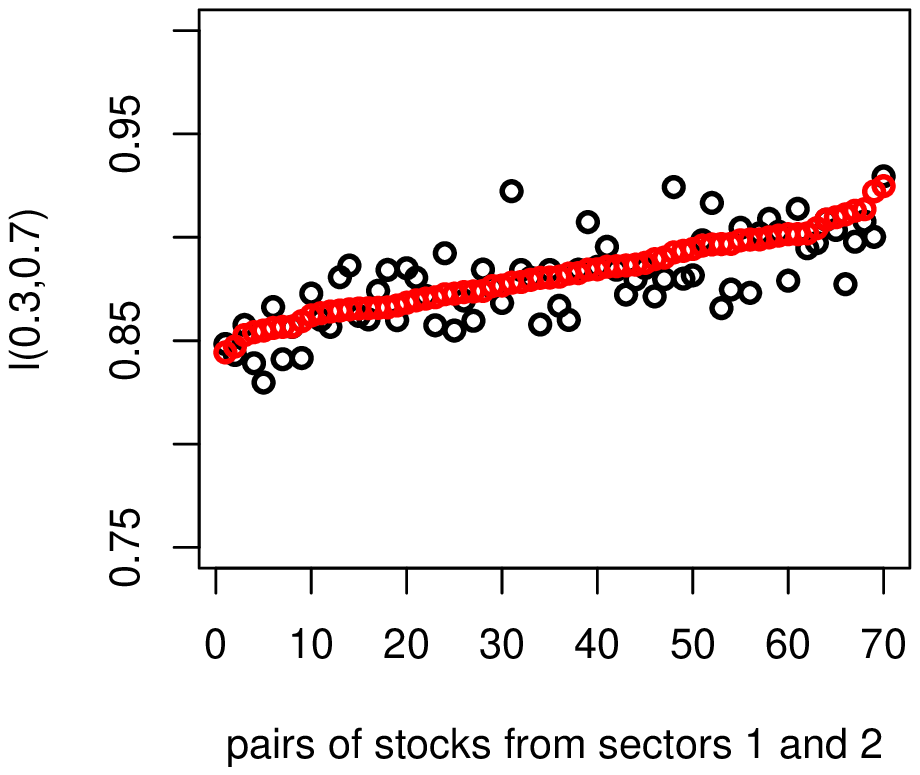}%\vspace{-0.6cm}
		\includegraphics[width=1.8in,height=1.9in]{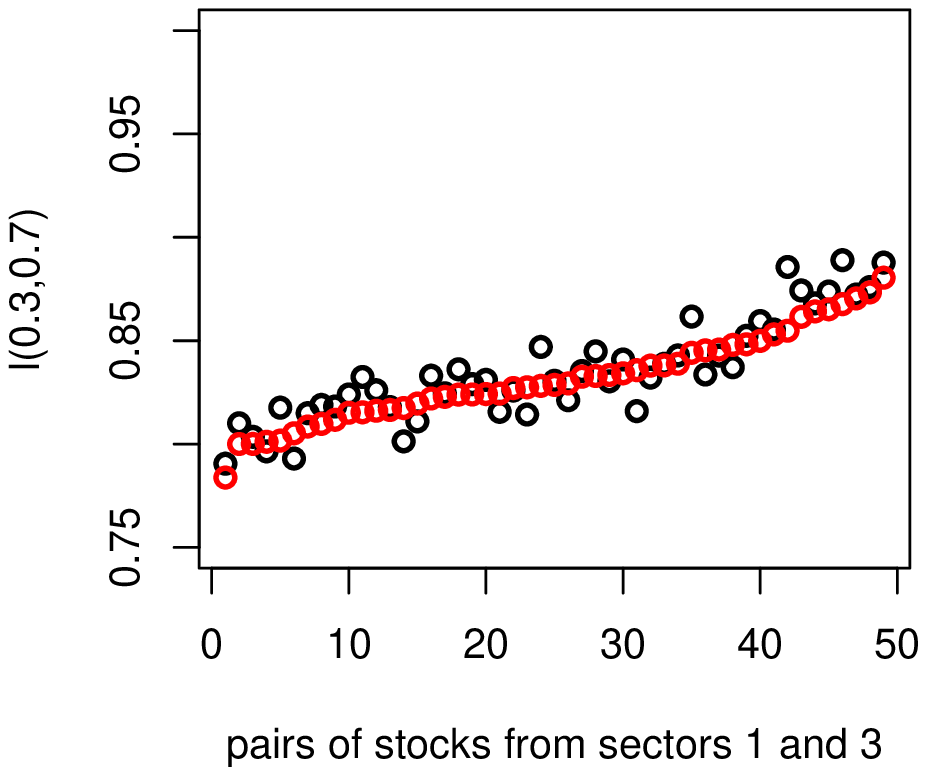}%\vspace{-0.6cm}
		\includegraphics[width=1.8in,height=1.9in]{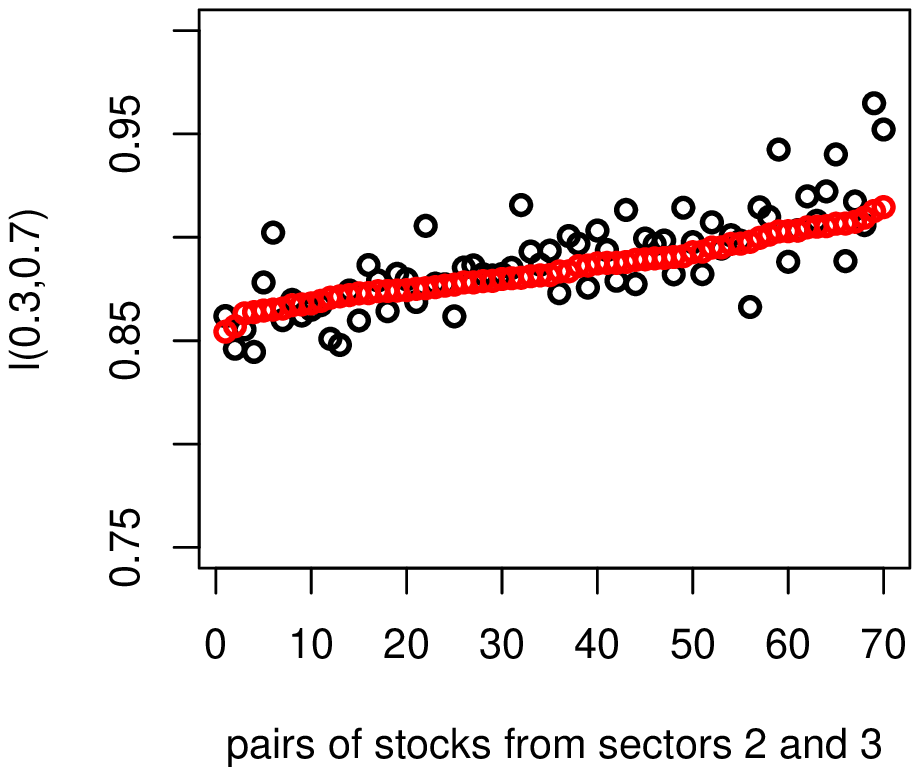}%\vspace{-0.6cm}
		\caption{{\footnotesize Empirical and the estimated tail dependence functions, $\ell(w, 1-w)$ (black and red points), for different pairs of stock returns  from sector 1 (top left), sector 2 (top middle), sector 3 (top right), and sectors 1, 2 (bottom left), sectors 1, 3 (bottom middle), sectors 2, 3 (bottom right) }}
		\label{fig_1bfct}
	\end{center}
\end{figure}

Similar to the wind data, we also calculate the empirical estimates of the tail dependence function $\ell(w, 1-w)$ for different pairs of stock returns, and we compare them against the model-based estimates. Figure \ref{fig_1bfct} shows the results for different pairs of stocks and $w = 0.3, 0.4$; we also obtained similar results for $w = 0.6, 0.7$. The results indicate that the proposed copula model with factor structure fits the data quite well.

We also check if the selected model is robust to linking copulas misspecification. We generate a sample of size $n=240$ from an extreme-value copula (\ref{eq-evcop}) with the Gumbel linking copulas. To replicate the dependence structure of the wind data, we use the following parameters
$$
\tht_{1:7} = 2.0, \quad \tht_{8:17} = 1.4, \quad \tht_{18:24} = 1.8, \quad \boldsymbol{\rho}_{1:7} = 0.5, \quad \boldsymbol{\rho}_{8:17} = 0.9, \quad \boldsymbol{\rho}_{18:24} = 0.9. 
$$  

Similar to the wind data, with these parameters, the overall strength of dependence for different pairs of variables (as measured by $S_{\rho}$) is close to the empirical correlations for the original data set. 
We then use model (\ref{eq-evcop}) with the misspecified reflected Clayton copulas and the correlation matrix $\Sig$ with a block-diagonal structure as before. For the estimated model, we compute the empirical and estimated values of $S_{\rho}$ and $\ell(w, 1-w)$ for $w=0.3, 0.4, 0.5$; Table \ref{tab5} reports the averaged values for pairs of variables from different sectors.

\begin{table}[h!]
	\caption{{\footnotesize Empirical / the estimated correlations $S_{\rho}$ and tail dependence functions $\ell(w,1-w)$, averaged for pairs of variables from different sectors. }}
	\label{tab5}
	\begin{center}
		{\small\begin{tabular}{l|cccccc}
				\hline
				sectors &1&2&3&1 and 2&1 and 3&2 and 3\\
				\hline
				%$S_{\rho}$&0.60/0.57&0.62/0.71&0.77/0.78&0.22/0.23&0.41/0.38&0.15/0.18\\
				%$\ell(0.7,0.3)$&0.70/0.70&0.72/0.69&0.64/0.64&0.92/0.92&0.85/0.86&0.95/0.93\\
				%$\ell(0.6,0.4)$&0.68/0.68&0.69/0.66&0.61/0.60&0.91/0.91&0.83/0.84&0.94/0.93\\
				%$\ell(0.5,0.5)$&0.67/0.68&0.69/0.65&0.59/0.59&0.91/0.91&0.83/0.84&0.94/0.92\\
				%$\ell(0.4,0.6)$&0.68/0.68&0.70/0.66&0.61/0.60&0.91/0.91&0.84/0.84&0.94/0.93\\
				%$\ell(0.3,0.7)$&0.70/0.70&0.72/0.69&0.64/0.64&0.92/0.92&0.86/0.86&0.94/0.93\\
				$S_{\rho}$&0.64/0.63&0.71/0.75&0.82/0.82&0.34/0.35&0.48/0.47&0.31/0.31\\
				$\ell(0.7,0.3)$&0.69/0.68&0.69/0.68&0.63/0.63&0.87/0.87&0.82/0.83&0.90/0.88\\
				$\ell(0.6,0.4)$&0.66/0.66&0.66/0.64&0.59/0.58&0.86/0.86&0.80/0.80&0.88/0.87\\
				$\ell(0.5,0.5)$&0.65/0.65&0.64/0.63&0.57/0.57&0.85/0.85&0.80/0.80&0.88/0.87\\
				$\ell(0.4,0.6)$&0.66/0.66&0.65/0.64&0.59/0.58&0.86/0.86&0.81/0.80&0.88/0.87\\
				$\ell(0.3,0.7)$&0.69/0.68&0.69/0.68&0.63/0.63&0.88/0.87&0.83/0.83&0.89/0.88\\
				\hline			
		\end{tabular}}
	\end{center}
\end{table}

Again, the empirical and model-based estimates are very close, and this indicates that the misspecified model has a reasonably good fit to the simulated data with a factor structure.

\section{Discussion}
\label{sec_disc}

In this paper, we considered a class of extreme-value copula models that are extreme-value limits of factor copula models with residual dependence modeled by a Gaussian copula. These are flexible models for multivariate extremes with complex dependence structures, such as spatial extremes or multivariate extremes with factor structures. Parsimonious dependence structures can be obtained with the  appropriate bivariate linking copulas and the correlation matrix $\Sig$ of the Gaussian copula. These models are computationally feasible in high dimensions, as only one-dimensional integration is required to compute the bivariate copula density. 

We used reflected Clayton linking copulas in the empirical study, but different linking copulas can be used to increase the flexibility of the proposed class of models. Also, the underlying dependence structure is assumed to be known (e.g., the partition of financial log-returns into groups/sectors). However, the underlying dependence structure is not always known, so model selection procedures and goodness-of-fit tests are topics requiring future research. Another topic for further research is to propose efficient estimation methods for this class of models involving higher-dimensional marginals and some computationally tractable special cases.   

%\section*{Acknowledgments}
%
%We would like to thank the associate editor and two anonymous referees for their constructive comments that helped to improve this paper. 

%\newpage
\setcounter{section}{1}
\renewcommand{\thesection}{\Alph{section}}
\section*{Appendix}

\subsection{Proof of Proposition \ref{prop3}}
\label{sec-appx0a}

With $\gamma_j = \frac{1}{k+1}A_j'(\frac{k}{k+1}) + A_j(\frac{k}{k+1})$ and $\eta_j(k) = (k+1)A_j(\frac{k}{k+1}) - k$, we find that $C_{j|0}(u|u^k) = \gamma_j u^{\eta_j(k)}$, $j=1,\ldots,d$, and
\begin{eqnarray*}
	C_{\UU}(\mathbf{u}) &=& \int_0^{\infty} C_N\{C_{1|0}(u|u^k), \ldots, C_{d|0}(u|u^k)\} u^k \ln u\,\df k \\
	&\leq& \int_0^{\infty} C_N\{\gamma^* u^{\eta^*(k)}, \ldots, \gamma^* u^{\eta^*(k)}\} u^k \ln u\,\df k \ \sim_{u \to 0} \ \int_0^{\infty} (\gamma^+u)^{\kappa_{\Sig}\cdot\eta^*(k) + k} \ell_N(u)\,\df k,
\end{eqnarray*}
where $\ell_N(u)$ is a slowly varying function and
$\gamma^* = \max_j\gamma_j,  \ \eta^*(k) = \min_j\eta_j(k)$. 
It is seen that $\kappa_L \geq \min_{k \geq 0}\{\kappa_{\Sig}\cdot\eta^*(k) + k\} > 1$ because $\eta^*(k) \geq \max(0, 1 - k)$ and $\eta^*(1) = \min_j \kappa_j - 1 > 0$.

Similarly, one can show that $\kappa_L \leq \min_{k \geq 0}\{\kappa_{\Sig}\cdot\eta^{**}(k) + k\}  \leq \kappa_{\Sig}\cdot\eta^{**}(0) = \kappa_{\Sig}$, where $\eta^{**}(k) = \max_j\eta_j(k)$. This implies that $C_{\UU}$ has intermediate lower tail dependence. \hfill $\Box$

\subsection{Proof of Proposition \ref{prop4}}
\label{sec-appx0b1}

Let $\bar C_{j|0}(u_j|u_0) = 1-C_{j|0}(1-u_j|1-u_0)$, $j=1,\ldots,d$. %Similar to Proposition \ref{prop1}, 
We use Theorem 8.76 of \cite{Joe2014}:  
\begin{eqnarray*}
	\ell(w_1,\ldots,w_d) &=& \lim_{u \to 0} \frac{1}{u}\left[1-\int_0^1C_N\{1-\bar C_{1|0}(uw_1|w_0),\ldots,1-\bar C_{d|0}(uw_d|w_0); \Sig\} \df w_0\right]\\
	&=& \lim_{u \to 0} \int_0^{1/u}\left[1 - C_N\{1-\bar C_{1|0}(uw_1|uw_0),\ldots,1-\bar C_{d|0}(uw_d|uw_0); \Sig\} \right]\df w_0\\
	&=& \int_0^{\infty}\left[1- C_N\{1-b_{1|0}(w_1|w_0),\ldots,1-b_{d|0}(w_d|w_0); \Sig\} \right]\df w_0. \quad\quad\quad\quad\quad  \Box
\end{eqnarray*}	
	
\subsection{Proof of Proposition \ref{prop5}}
\label{sec-appx0b2}

We have $C_{j|0}(1-uw_j|v_0) = 1-uw_jc_{j,0}(1,v_0) + o(u)$, $j=1,\ldots,d$, and therefore
\begin{eqnarray*}
	\ell(w_1, \ldots, w_d) &=& \lim_{u\to 0}\frac{1}{u}\int_0^1\left[1 - \min_j\{C_{j|0}(1-uw_j| v_0)\}\right]\df v_0\\
	&=&    \lim_{u\to 0}\frac{1}{u}\int_0^1\left[uw_j\max_j\{c_{j,0}(1, v_0) + o(1)\}\right]\df v_0 = \int_0^1 \max_j\{w_jc_{j,0}(1,v_0)\} \df v_0. \quad\quad\Box
\end{eqnarray*}

\subsection{Proof of Proposition \ref{prop6}}
\label{sec-appx0}

We  use the following Lemma to prove this proposition.

\medskip
\emph{Lemma 1:} Let $C_N(u,v; \rho)$ be the normal copula with correlation $\rho$. If $\rho \to 1$, then $$C_N(u,u;\rho) = u - \left(\frac{1-\rho}{\pi}\right)^{1/2}\phi\{\Phi^{-1}(u)\} + (1-\rho)^{3/2}\phi\{\Phi^{-1}(u)\}[K_1(\rho) + K_2(\rho)\{\Phi^{-1}(u)\}^2],$$ where  $\max_{\rho} |K_i(\rho)| \leq K_0 < \infty$, $i=1,2$.

\emph{Proof of Lemma 1}: Denote $C_N(u|v; \rho) = \p C_N(u, v; \rho)/ \p v$ and $\delta = \sqrt{1-\rho}$. We find that $$C_N(u,u;\rho) = 2\int_0^u C_N(v|v;\rho)\df v =  2\int_0^u\Phi\left\{\frac{\delta}{\sqrt{2-\delta^2}}\Phi^{-1}(v)\right\} \df v.$$
Let $h(\delta) = h(\delta, v) = \Phi\left\{\frac{\delta}{\sqrt{2-\delta^2}}\Phi^{-1}(v)\right\}$ and $g(\delta) = g(\delta, v) =  \phi\left\{\frac{\delta}{\sqrt{2-\delta^2}}\Phi^{-1}(v)\right\}$. Using the Taylor expansion of $h(\delta)$ around $\delta = 0$ with a fixed $v$ yields:
$$
h(\delta) = h(0) + h'(0)\delta + h''(\varUpsilon)\delta^2, \quad 0 < \varUpsilon < \delta,
$$
where 
$$
h'(t) = \frac{2\Phi^{-1}(v)}{(2-t^2)^{1.5}} \cdot g(t), \quad h''(t) = -\frac{2t\Phi^{-1}(v)}{(2-t^2)^{3.5}} \cdot [2\{\Phi^{-1}(v)\}^2 - 3(2-t^2)]g(t).
$$
It implies that $$h(\delta) = 0.5 + 0.5\pi^{-1/2}\delta \Phi^{-1}(v) + w_1\delta^3 \Phi^{-1}(v) - w_2\delta^3 \{\Phi^{-1}(v)\}^3,$$
where $0 < w_1 < 6\phi(0)$ and $0 < w_2 < 4\phi(0)$ for any $0 < v < 1$ and $\delta \to 0$; hence,
$$
C_N(u,u;\rho) = 2\int_0^u h(\delta, v) \df v = u + \pi^{-1/2}\delta I_1 + 2w_1\delta^3 I_1 - 2w_2\delta^3 I_2, 
$$
where
$$
I_1 = \int_0^u \Phi^{-1}(v) \df v = -\phi\{\Phi^{-1}(u)\}, \quad I_2 = \int_0^u \{\Phi^{-1}(v)\}^3 \df v = -[2+\{\Phi^{-1}(u)\}^2]\phi\{\Phi^{-1}(u)\}.
$$
Finally,
$$
C_N(u,u;\rho) = u - \pi^{-1/2}\delta \phi\{\Phi^{-1}(u)\} + \delta^3[K_1 + K_2\{\Phi^{-1}(u)\}^2]\phi\{\Phi^{-1}(u)\},
$$
where $K_1 = -2w_1 + 4w_2$, $K_2 = 2w_2$ and $|K_1| < 24\phi(0)$, $K_2 < 12\phi(0)$. \hfill $\Box$

\medskip
\emph{Proof o Proposition \ref{prop6}}: Note that $\phi\{\Phi^{-1}(u)\} \sim u \ell^*(u)$ as $u \to 0$, where $\ell^*(u)$ is a slowly varying function. It implies that for any $m \geq 0$,
$$\int_0^{\infty} [\Phi^{-1}\{b_{1|0}(1|w_0)\}]^m\phi[\Phi^{-1}\{b_{1|0}(1|w_0)\}] \df w_0 < \infty.$$ From Proposition \ref{prop4} and Lemma 1, we find that, as $\rho \to 1$,
\begin{eqnarray*}
	\ell(1,1) &=& \int_0^{\infty}[1- C_N\{1-b_{1|0}(1|w_0), 1-b_{1|0}(1|w_0); \rho\}] \df w_0\\
	&=& 2-\int_0^{\infty}C_N\{b_{1|0}(1|w_0), b_{1|0}(1|w_0); \rho\} \df w_0\\
	&=& 2 - \int_0^{\infty}b_{1|0}(1|w_0) \df w_0 + \left(\frac{1-\rho}{\pi}\right)^{1/2}\int_0^{\infty} \phi[\Phi^{-1}\{b_{1|0}(1|w_0)\}] \df w_0 + O((1-\rho)^{3/2})\\
	&=& 1 + \left(\frac{1-\rho}{\pi}\right)^{1/2}\int_0^{\infty} \phi[\Phi^{-1}\{b_{1|0}(1|w_0)\}] \df w_0 + O((1-\rho)^{3/2}). \hspace{3.2cm} \Box
\end{eqnarray*}

\subsection{Computation of $V_{j,k}(w_j, w_k)$}
\label{sec-appx1}

Let $I_{j,k}(w_0) = 1 - C_N\{1-b_{j|0}(w_j|w_0),1-b_{k|0}(w_k|w_0); \rho_{j,k} \}$. We assume that, as $w_0 \to \infty$, $b_{j|0}(w_j|w_0) = \ell_j(w_0) w_0^{-\phi_j} + o(w_0^{-\phi_j})$ and $b_{k|0}(w_k|w_0) = \ell_k(w_0) w_0^{-\phi_k} + o(w_0^{-\phi_k})$, $\phi_j, \phi_k > 1$, where $\ell_j$ and $\ell_k$ are slowly varying functions. It follows that $I_{j,k}(w_0) = \ell_{j,k}(w_0)w_0^{-\phi_{j,k}}$ as $w_0 \to \infty$ where $\phi_{j,k} = \min(\phi_j, \phi_k)$ and $\ell_{j,k}$ is a slowly varying function. The integrand $I_{j,k}(w_0)$ has a slow rate of decay in the tail and standard numerical integration methods used to compute $V_{j,k}(w_j,w_k) = \int_0^{\infty} I_{j,k}(w_0) \df w_0$ may not be efficient. 

To make computations more efficient, we can write
$$
V_{j,k}(w_j,w_k) = \int_0^1 I_{j,k}(w_0) \df w_0 + \int_1^{\infty} I_{j,k}(w_0) \df w_0 =  \int_0^1 I_{j,k}(w_0) \df w_0 + \alpha\int_0^{1} I_{j,k}(z_0^{-\alpha}) z_0^{-\alpha - 1}\df z_0,
$$
where the first integral has finite integration limits and bounded integrand. The second integrand can take large values if $z_0$ is close to zero. We therefore need to select the smallest $\alpha > 0$ such that $I_{j,k}^*(z_0) = I_{j,k}(z_0^{-\alpha})z_0^{-\alpha - 1} < \infty$ for $0 \leq z_0 \le 1$. We have $I^*_{j,k}(z_0) = \ell_{j,k}(z_0^{-\alpha})z_0^{(\phi_{j,k}-1)\alpha-1}$ as $z_0 \to 0$ and one can select $\alpha = \alpha_{j,k} = \{\phi_{j,k}-1\}^{-1}$. Now Gauss-Legendre quadrature \citep{Stroud.Secrest1966} can be used to evaluate the two integrals and compute $V_{j,k}(w_j, w_k)$.

The assumption about the tail behavior of $b_{j|0}$ holds for many copulas with  the upper tail dependence. For the reflected Clayton copula with parameter $\theta_j$, $$b_{j|0}(w_j|w_0) = \left\{1+(w_0/w_j)^{\theta_j}\right\}^{-1-1/\theta_j} = (w_0/w_j)^{-1-\theta_j} + o(w_0^{-1-\theta_j}), \quad \text{as } w_0 \to \infty,$$ and therefore $\alpha_{j,k} = 1/\min(\theta_j, \theta_k)$. For the Gumbel copula with parameter $\theta_j$, $$b_{j|0}(w_j|w_0) = 1-\left\{1+(w_j/w_0)^{\theta_j}\right\}^{-1+1/\theta_j} = (1-1/\theta_j)(w_0/w_j)^{-\theta_j} + o(w_0^{-\theta_j}), \quad \text{as } w_0 \to \infty,$$ and therefore $\alpha_{j,k} = 1/\{\min(\theta_j, \theta_k)-1\}$.

Similar ideas can be used to compute the derivatives of the stable tail dependence function. We found that the same transformation works very well in this case and that this change of variables greatly improves the accuracy of numerical integration and $n_q = 35$ quadrature points is sufficient to compute the density $c_{j,k}^{\EV}$ in most cases. 

\subsection{Parameter estimation for $C_{\UU}^{\EV}$ with Clayton linking copulas}
\label{sec-appx2}

Here we provide more details for $C_{\UU}^{\EV}$ in Section \ref{subsec-mm-est} with Clayton linking copulas. Note that $V_{j,k}(1,1) = 1$ if $\theta_j = \theta_k$. Without loss of generality, we now assume that $\theta_j > \theta_k$ for the $(j,k)$ margin. We have $c(1,v;\theta) = (\theta+1)v^{\theta}$ and $$V_{j,k}(1,1) = \int_0^{v_{j,k}} c(1,v_0;\theta_k)\df v_0 + \int_{v_{j,k}}^1 c(1,v_0;\theta_j)\df v_0 = 1 - C(v_{j,k}|1; \theta_j) + C(v_{j,k}|1; \theta_k),$$
where the conditional Clayton copula $C(v|1; \theta) = v^{\theta+1}$ and $v_{j,k} \in (0,1)$ satisfies   
$$
c(1,v_{j,k}; \theta_j) = c(1,v_{j,k}; \theta_k) \quad \Rightarrow \quad v_{j,k} = \left(\frac{\theta_k+1}{\theta_j+1}\right)^{\frac{1}{\theta_j - \theta_k}}\,,
$$ 
and therefore $$V_{j,k}(1,1) = 1 + \frac{(\vartheta-1)^{\vartheta-1}}{\vartheta^{\vartheta}}, \quad \vartheta = \frac{\theta_j+1}{\theta_j - \theta_k}\,.$$

Similarly, one can show that the copula $C_{\UU}^{\EV}$ and its lower-dimensional marginals depend on $\vartheta(\theta_j, \theta_k) = (\theta_j+1)/(\theta_j - \theta_k)$, or, equivalently, on $\vartheta^*(\theta_j, \theta_k) = \{\vartheta(\theta_j,\theta_k)-1\}/\vartheta(\theta_j,\theta_k) = (\theta_k+1)/(\theta_j+1)$ for $1 \leq k < j \leq d$. The model is therefore non-identifiable and one can fix one parameter and estimate the remaining $d-1$ parameters. 

If the order of variables is ignored, the model is still non-identifiable even if one parameter is fixed.

\medskip
\textbf{Example 5:} Assume that $d=3$ and $\tht = (0.5, 1, 2)^{\top}$. We generate a sample of size $N=100$ from $C_{\UU}^{\EV}$ assuming $\Sig$ is a matrix of ones. We fix $\tht_2=1$ and find that the objective function (\ref{eq-argmin}) attains its minimum at $\widehat\tht = (0.456, 1, 2.584)^{\top}$ and $\tilde\tht = (1.747, 1, 0.116)^{\top}$. We can see that $$\frac{\widehat\tht_1+1}{\widehat\tht_2+1} = \frac{\tilde\tht_2+1}{\tilde\tht_1+1}\,, \quad \frac{\widehat\tht_1+1}{\widehat\tht_3+1} = \frac{\tilde\tht_3+1}{\tilde\tht_1+1}\,, \quad  \frac{\widehat\tht_2+1}{\widehat\tht_3+1} = \frac{\tilde\tht_3+1}{\tilde\tht_2+1}\,.$$

\begin{figure}[h!]
	\begin{center}
		\includegraphics[width=2.1in,height=2.35in]{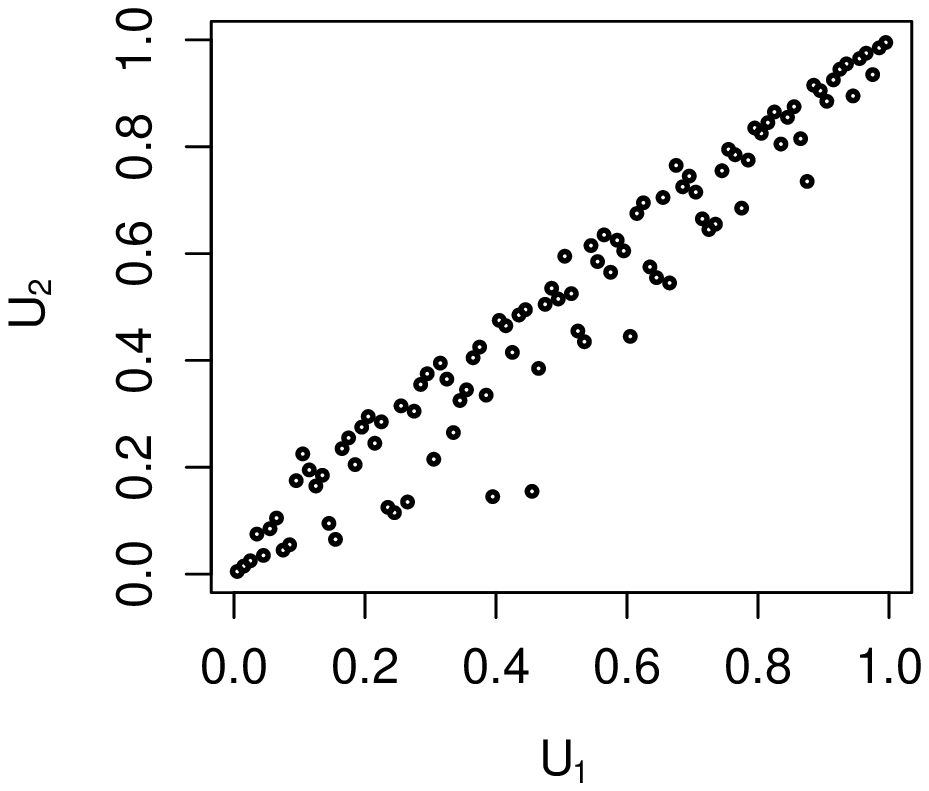}
		\includegraphics[width=2.1in,height=2.35in]{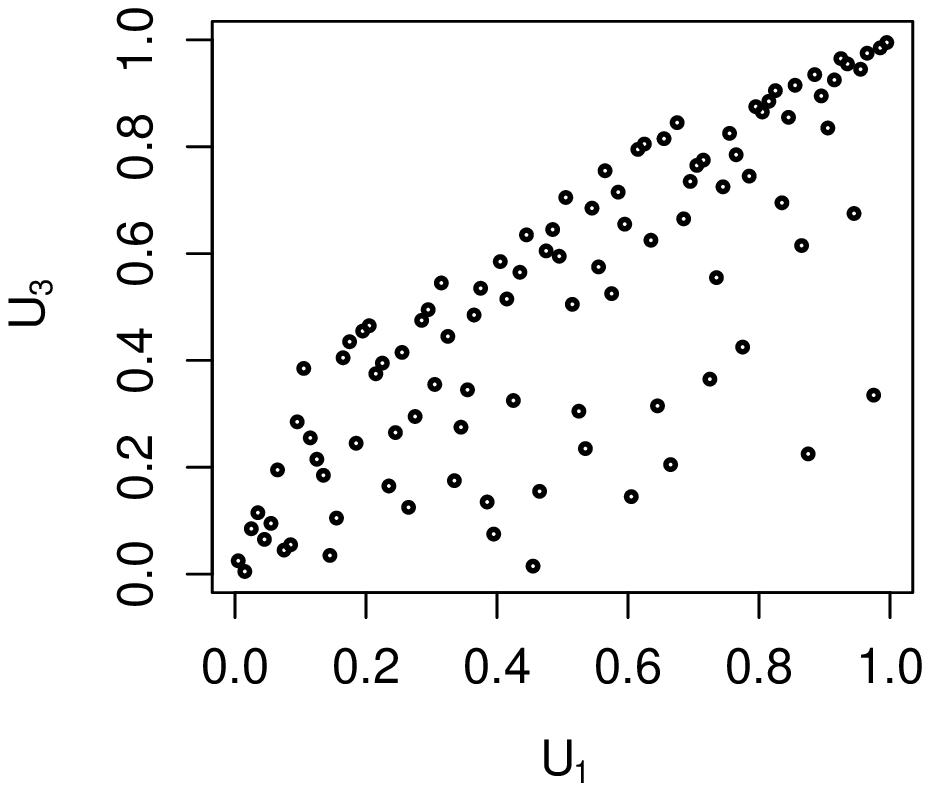}
		\includegraphics[width=2.1in,height=2.35in]{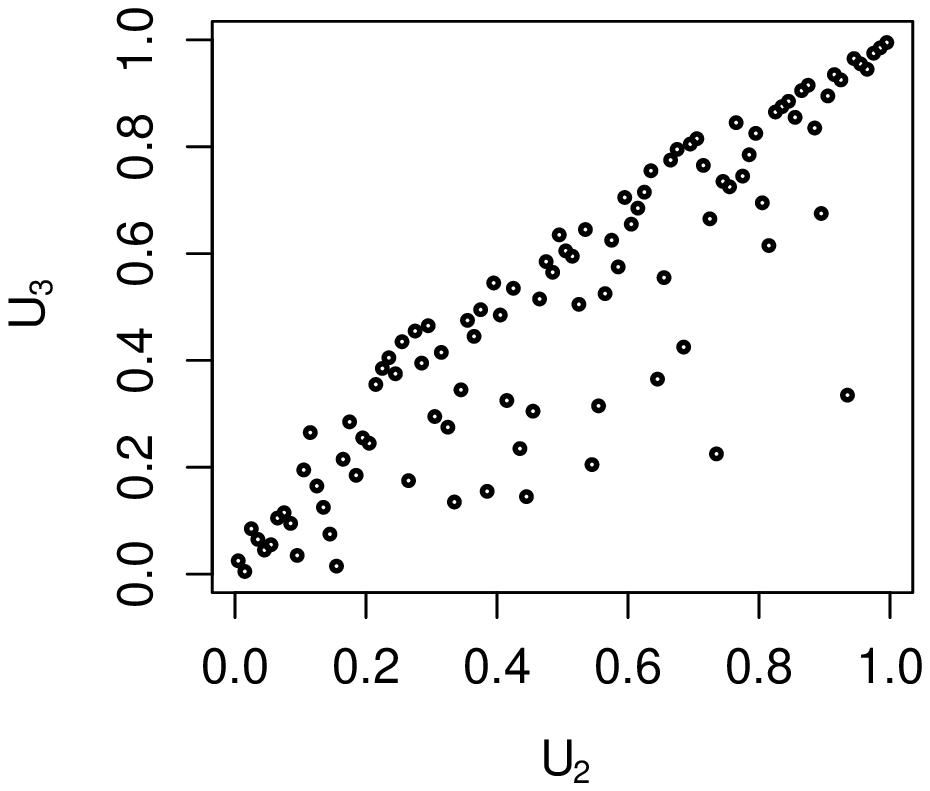}
		\caption{{\footnotesize Bivariate scatter plots for the data set simulated from $C_{\UU}^{\EV}$ with degenerate $\Sig$ (matrix of ones) and Clayton linking copulas with $\tht=(0.5,1,2)^{\top}$ }}
		\label{figCCC}
	\end{center}
\end{figure}

To select the right solution, one can check bivariate scatter plots: if $\theta_j > \theta_k$ for the $(j,k)$ margin, the marginal density is zero around the corner $(0,1)$ and the density is skewed to the lower right corner. Figure \ref{figCCC} shows scatter plots for the simulated data set. The plots indicate that $\theta_1 < \theta_2 < \theta_3$ and therefore $\widehat\tht$ should be selected.

\bibliographystyle{model2-names}

\begin{thebibliography}{26}
	\expandafter\ifx\csname natexlab\endcsname\relax\def\natexlab#1{#1}\fi
	\expandafter\ifx\csname url\endcsname\relax
	\def\url#1{\texttt{#1}}\fi
	\expandafter\ifx\csname urlprefix\endcsname\relax\def\urlprefix{URL }\fi
	\providecommand{\eprint}[2][]{\url{#2}}
	\providecommand{\bibinfo}[2]{#2}
	\ifx\xfnm\relax \def\xfnm[#1]{\unskip,\space#1}\fi
	%Type = Article
	\bibitem[{Aas et~al.(2009)Aas, Czado, Frigessi and Bakken}]{Aas.Czado.ea2009}
	\bibinfo{author}{Aas, K.}, \bibinfo{author}{Czado, C.},
	\bibinfo{author}{Frigessi, A.}, \bibinfo{author}{Bakken, H.},
	\bibinfo{year}{2009}.
	\newblock \bibinfo{title}{Pair-copula constructions of multiple dependence}.
	\newblock \bibinfo{journal}{Insurance: Mathematics and Economics}
	\bibinfo{volume}{44}, \bibinfo{pages}{182--198}.
    %Type = Article
	\bibitem[{Aloui et~al.(2014)Aloui, Aissa, Hammoudeh and Nguyen}]{Aloui.Aissa.ea2014}
	\bibinfo{author}{Aloui, R.}, \bibinfo{author}{A\"{i}ssa, M.S.B.},
	\bibinfo{author}{Hammoudeh, S.}, \bibinfo{author}{Nguyen, D.K.},
	\bibinfo{year}{2009}.
	\newblock \bibinfo{title}{Dependence and extreme dependence of crude oil and natural gas prices with applications to risk management}.
	\newblock \bibinfo{journal}{Energy Economics}
	\bibinfo{volume}{42}, \bibinfo{pages}{332--342}.
	
    %Type = Article
	\bibitem[{Andersen(2004)}]{Andersen2004}
	\bibinfo{author}{Andersen, E.\,W.}, \bibinfo{year}{2004}.
	\newblock \bibinfo{title}{Composite likelihood and two-stage estimation in
		family studies}.
	\newblock \bibinfo{journal}{Biostatistics}
	\bibinfo{volume}{5(1)}, \bibinfo{pages}{15--30}.
	
	%Type = Article
	\bibitem[{Bargaoui and Bardossy(2015)}]{Bargaoui.Bardossy.2015}
	\bibinfo{author}{Bargaoui, Z.K.}, \bibinfo{author}{Bardossy, A.},
	\bibinfo{year}{2015}.
	\newblock \bibinfo{title}{Modeling short duration extreme precipitation patterns using copula and generalized maximum pseudo-likelihood estimation with censoring}.
	\newblock \bibinfo{journal}{Advances in Water Resources}
	\bibinfo{volume}{84},
	\bibinfo{pages}{1--13}.
	%Type = Article
	\bibitem[{Brechmann et~al.(2012)Brechmann, Czado and
		Aas}]{Brechmann.Czado.ea2012}
	\bibinfo{author}{Brechmann, E.C.}, \bibinfo{author}{Czado, C.},
	\bibinfo{author}{Aas, K.}, \bibinfo{year}{2012}.
	\newblock \bibinfo{title}{Truncated regular vines in high dimensions with
		applications to financial data}.
	\newblock \bibinfo{journal}{Canadian Journal of Statistics}
	\bibinfo{volume}{40}, \bibinfo{pages}{68--85}.
	
	%Type = Article
	\bibitem[{Castro and Huser(2020)}]{Castro.Huser.2020}
	\bibinfo{author}{Castro-Camilo, D.}, \bibinfo{author}{Huser, R.}, \bibinfo{year}{2020}.
	\newblock \bibinfo{title}{ Local likelihood estimation of complex tail dependence structures, applied to U.S. precipitation extremes}.
	\newblock \bibinfo{journal}{Journal of the American Statistical Association} \bibinfo{volume}{115},
	\bibinfo{pages}{1037--1045}.
	
	%Type = Article
	\bibitem[{Demarta and McNeil(2005)}]{Demarta.McNeil.2005}
	\bibinfo{author}{Demarta, S.}, \bibinfo{author}{McNeil, A.J.},
	\bibinfo{year}{2005}.
	\newblock \bibinfo{title}{The t copula and related copulas}.
	\newblock \bibinfo{journal}{International Statistical Review} \bibinfo{volume}{73},
	\bibinfo{pages}{111--129}.
	%Type = Article
	\bibitem[{Ferreira(2013)}]{Ferreira2013}
	\bibinfo{author}{Ferreira, M.}, \bibinfo{year}{2013}.
	\newblock \bibinfo{title}{Nonparametric estimation of the tail-dependence
		coefficient}.
	\newblock \bibinfo{journal}{REVSTAT -- Statistical Journal}
	\bibinfo{volume}{11}, \bibinfo{pages}{1--16}.
	%Type = Article
	\bibitem[{Genest et~al.(1995)Genest, Ghoudi and Rivest}]{Genest.Ghoudi.ea1995}
	\bibinfo{author}{Genest, C.}, \bibinfo{author}{Ghoudi, K.},
	\bibinfo{author}{Rivest, L.P.}, \bibinfo{year}{1995}.
	\newblock \bibinfo{title}{A semiparametric estimation procedure of dependence
		parameters in multivariate families of distributions}.
	\newblock \bibinfo{journal}{Biometrika} \bibinfo{volume}{82},
	\bibinfo{pages}{543--552}.
	%Type = Article
	\bibitem[{Genest and Segers(2009)}]{Genest.Segers2009}
	\bibinfo{author}{Genest, C.}, \bibinfo{author}{Segers, J.},
	\bibinfo{year}{2009}.
	\newblock \bibinfo{title}{Rank-based inference for bivariate extreme-value
		copulas}.
	\newblock \bibinfo{journal}{Annals of Statistics} \bibinfo{volume}{37},
	\bibinfo{pages}{2990--3022}.
	%Type = Article
	\bibitem[{Genton and Kleiber(2015)}]{Genton.Kleiber2015}
	\bibinfo{author}{Genton, M.\, G.}, \bibinfo{author}{Kleiber, W.},
	\bibinfo{year}{2015}.
	\newblock \bibinfo{title}{Cross-covariance functions for multivariate
		geostatistics (with discussion)}.
	\newblock \bibinfo{journal}{Statistical Science} \bibinfo{volume}{30},
	\bibinfo{pages}{147--163}.
	%Type = Book
	\bibitem[{Gneiting et~al.(2007)Gneiting, Genton and
		Guttorp}]{Gneiting.Genton.ea2007}
	\bibinfo{author}{Gneiting, T.}, \bibinfo{author}{Genton, M.\, G.},
	\bibinfo{author}{Guttorp, P.}, \bibinfo{year}{2007}.
	\newblock \bibinfo{title}{Geostatistical space-time models, stationarity,
		separability and full symmetry}.
	\newblock \bibinfo{publisher}{In Finkenstaedt, B., Held, L. and Isham, V.(eds),
		{\it Statistics of Spatio-Temporal Systems}, Chapman \& Hall / CRC Press,
		Monograph in Statistics and Applied Probability}, \bibinfo{address}{Boca
		Raton}.
	%Type = Book
	\bibitem[{de~Haan and Ferreira(2006)}]{deHaan.Ferreira.2006}
	\bibinfo{author}{de~Haan, L.}, \bibinfo{author}{Ferreira, A.},
	\bibinfo{year}{2006}.
	\newblock \bibinfo{title}{Extreme Value Theory}.
	\newblock \bibinfo{publisher}{Springer-Verlag}, \bibinfo{address}{New York}.
	%Type = Article
	\bibitem[{Hua and Joe(2011)}]{Hua.Joe2011}
	\bibinfo{author}{Hua, L.}, \bibinfo{author}{Joe, H.}, \bibinfo{year}{2011}.
	\newblock \bibinfo{title}{Tail order and intermediate tail dependence of
		multivariate copulas}.
	\newblock \bibinfo{journal}{Journal of Multivariate Analysis}
	\bibinfo{volume}{102}, \bibinfo{pages}{1454--1471}.
	
    %Type = Article
	\bibitem[{Hua et~al.(2017)Hua, Xia and
		Basu}]{Hua.Xia.ea2017}
	\bibinfo{author}{Hua, L.}, \bibinfo{author}{Xia, M.},
	\bibinfo{author}{Basu, S.}, \bibinfo{year}{2017}.
	\newblock \bibinfo{title}{Factor copula approaches for assessing spatially
		dependent high-dimensional risks}.
	\newblock \bibinfo{journal}{North American Actuarial Journal}
	\bibinfo{volume}{21(1)}, \bibinfo{pages}{147--160}.
	
	%Type = Article
	\bibitem[{H\"{u}sler and Reiss(1989)}]{Husler.Reiss1989}
	\bibinfo{author}{H\"{u}sler, J.}, \bibinfo{author}{Reiss, R.\, D.},
	\bibinfo{year}{1989}.
	\newblock \bibinfo{title}{Maxima of normal random vectors: between independence
		and complete dependence}.
	\newblock \bibinfo{journal}{Statistics and Probability Letters}
	\bibinfo{volume}{7}, \bibinfo{pages}{283--286}.
	%Type = Book
	\bibitem[{Joe(2014)}]{Joe2014}
	\bibinfo{author}{Joe, H.}, \bibinfo{year}{2014}.
	\newblock \bibinfo{title}{Dependence Modeling with Copulas}.
	\newblock \bibinfo{publisher}{Chapman \& Hall/CRC}, \bibinfo{address}{Boca
		Raton, FL}.
	
    %Type = Article
	\bibitem[{Joe and Xu(1996)}]{Joe.Xu1996}
	\bibinfo{author}{Joe, H.}, \bibinfo{author}{Xu, J.\,J.},
	\bibinfo{year}{1996}.
	\newblock \bibinfo{title}{The estimation method of inference functions for margins for multivariate models}.
	\newblock \bibinfo{journal}{Technical Report} \bibinfo{volume}{166},
	%\bibinfo{pages}{2990--3022}.
	
	%Type = Article
	\bibitem[{Krupskii and Genton(2017)}]{Krupskii.Genton2017}
	\bibinfo{author}{Krupskii, P.}, \bibinfo{author}{Genton, M.\, G.},
	\bibinfo{year}{2017}.
	\newblock \bibinfo{title}{Factor copula models for data with spatio-temporal dependence}.
	\newblock \bibinfo{journal}{Spatial Statistics}
	\bibinfo{volume}{22}, \bibinfo{pages}{108--195}.
	
	%Type = Article
	\bibitem[{Krupskii and Genton(2018)}]{Krupskii.Genton2018}
	\bibinfo{author}{Krupskii, P.}, \bibinfo{author}{Genton, M.\, G.},
	\bibinfo{year}{2018}.
	\newblock \bibinfo{title}{Linear factor copula models and their properties}.
	\newblock \bibinfo{journal}{Scandinavian Journal of Statistics}
	\bibinfo{volume}{45(4)}, \bibinfo{pages}{861--878}.
	
	%Type = Article
	\bibitem[{Krupskii and Genton(2019)}]{Krupskii.Genton2019}
	\bibinfo{author}{Krupskii, P.}, \bibinfo{author}{Genton, M.\, G.},
	\bibinfo{year}{2019}.
	\newblock \bibinfo{title}{A copula model for non-Gaussian multivariate spatial data}.
	\newblock \bibinfo{journal}{Journal of Multivariate Analysis}
	\bibinfo{volume}{169}, \bibinfo{pages}{264--277}.
	
	%Type = Article
	\bibitem[{Krupskii et~al.(2018a)Krupskii, Huser and
		Genton}]{Krupskii.Huser.ea2016}
	\bibinfo{author}{Krupskii, P.}, \bibinfo{author}{Huser, R.},
	\bibinfo{author}{Genton, M.\, G.}, \bibinfo{year}{2018}a.
	\newblock \bibinfo{title}{Factor copula models for replicated spatial data}.
	\newblock \bibinfo{journal}{Journal of the American Statistical Association}
	\bibinfo{volume}{521}, \bibinfo{pages}{467--479}.
	%Type = Article
	\bibitem[{Krupskii and Joe(2013)}]{Krupskii.Joe2013}
	\bibinfo{author}{Krupskii, P.}, \bibinfo{author}{Joe, H.},
	\bibinfo{year}{2013}.
	\newblock \bibinfo{title}{Factor copula models for multivariate data}.
	\newblock \bibinfo{journal}{Journal of Multivariate Analysis}
	\bibinfo{volume}{120}, \bibinfo{pages}{85--101}.
	%Type = Article
	\bibitem[{Krupskii and Joe(2015)}]{Krupskii.Joe2015b}
	\bibinfo{author}{Krupskii, P.}, \bibinfo{author}{Joe, H.},
	\bibinfo{year}{2015}.
	\newblock \bibinfo{title}{Structured factor copula models: theory, inference
		and computation}.
	\newblock \bibinfo{journal}{Journal of Multivariate Analysis}
	\bibinfo{volume}{138}, \bibinfo{pages}{53--73}.
	%Type = Article
	\bibitem[{Krupskii and Joe(2020)}]{Krupskii.Joe2019}
	\bibinfo{author}{Krupskii, P.}, \bibinfo{author}{Joe, H.},
	\bibinfo{year}{2020}.
	\newblock \bibinfo{title}{Flexible copula models with dynamic dependence and
		application to financial data}.
	\newblock \bibinfo{journal}{Journal of Econometrics and Statistics.}
	\bibinfo{volume}{16}, \bibinfo{pages}{148--167}
	%Type = Article
	\bibitem[{Krupskii et~al.(2018b)Krupskii, Joe, Lee and
		Genton}]{Krupskii.Joe.ea2018}
	\bibinfo{author}{Krupskii, P.}, \bibinfo{author}{Joe, H.},
	\bibinfo{author}{Lee, D.}, \bibinfo{author}{Genton, M.\, G.},
	\bibinfo{year}{2018}b.
	\newblock \bibinfo{title}{Extreme value limit of the convolution of exponential
		and multivariate normal distributions: Link to the {H}\"usler-{R}eiss
		distribution}.
	\newblock \bibinfo{journal}{Journal of Multivariate Analysis}
	\bibinfo{volume}{163}, \bibinfo{pages}{80--95}.
    %Type = Book
	\bibitem[{Kurowicka and Cooke(2006)}]{Kurowicka.Cooke2006}
	\bibinfo{author}{Kurowicka, D.}, \bibinfo{author}{Cooke, R.},
	\bibinfo{year}{2006}.
	\newblock \bibinfo{title}{Uncertainty Analysis with High Dimensional Dependence
		Modelling}.
	\newblock \bibinfo{publisher}{Wiley Series in Probability and Statistics}.
	%Type = Article
	\bibitem[{Lee and Joe(2018)}]{Lee.Joe.2018}
	\bibinfo{author}{Lee, D.}, \bibinfo{author}{Joe, H.}, \bibinfo{year}{2018}.
	\newblock \bibinfo{title}{Multivariate extreme value copulas with factor and
		tree dependence structures}.
	\newblock \bibinfo{journal}{Extremes} \bibinfo{volume}{21(1)},
	\bibinfo{pages}{147--176}.
	%Type = Article
	\bibitem[{Lindsay(1998)}]{Lindsay.1998}
	\bibinfo{author}{Lindsay, B.}, \bibinfo{year}{1998}.
	\newblock \bibinfo{title}{Composite likelihood methods}.
	\newblock \bibinfo{journal}{Contemporary Mathematics} \bibinfo{volume}{80},
	\bibinfo{pages}{220--239}.
	%Type = Article
	\bibitem[{Manner and Segers(2011)}]{Manner.Segers.2011}
	\bibinfo{author}{Manner, H.}, \bibinfo{author}{Segers, J.}, \bibinfo{year}{2011}.
	\newblock \bibinfo{title}{Tails of correlation mixtures of elliptical copulas}.
	\newblock \bibinfo{journal}{Insurance: Mathematics and Economics} \bibinfo{volume}{48(1)},
	\bibinfo{pages}{153--160}.
	%Type = Book
	\bibitem[{McNeil et~al.(2005)McNeil, Frey and Embrechts}]{McNeil.Frey.ea2005}
	\bibinfo{author}{McNeil, A.\,J.}, \bibinfo{author}{Frey, R.},
	\bibinfo{author}{Embrechts, P.}, \bibinfo{year}{2005}.
	\newblock \bibinfo{title}{Quantitative Risk Management}.
	\newblock \bibinfo{publisher}{Princeton University Press},
	\bibinfo{address}{Princeton, NJ}.
	%Type = Article
	\bibitem[{Millar(1984)}]{Millar.1984}
	\bibinfo{author}{Millar, P.\,W.}, \bibinfo{year}{1984}.
	\newblock \bibinfo{title}{A general approach to the optimality of minimum distance estimators}.
	\newblock \bibinfo{journal}{Transactions of the American Mathematical Society} \bibinfo{volume}{286},
	\bibinfo{pages}{377--418}.
	
	%Type = Article
	\bibitem[{Newey and McFadden(1994)}]{Newey.McFadden1994}
	\bibinfo{author}{Newey, W.\, K.}, \bibinfo{author}{McFadden, D.},
	\bibinfo{year}{1994}.
	\newblock \bibinfo{title}{Large sample estimation and hypothesis testing}.
	\newblock \bibinfo{journal}{Handbook of Econometrics} \bibinfo{volume}{4}.
	%\bibinfo{pages}{2111--2245}.
	
	%Type = Article
	\bibitem[{Nikoloulopoulos et~al.(2009)Nikoloulopoulos, Joe and
		Li}]{Nikoloulopoulos.Joe.ea2009}
	\bibinfo{author}{Nikoloulopoulos, A.\, K.}, \bibinfo{author}{Joe, H.},
	\bibinfo{author}{Li, H.}, \bibinfo{year}{2009}.
	\newblock \bibinfo{title}{Extreme value properties of multivariate t copulas}.
	\newblock \bibinfo{journal}{Extremes} \bibinfo{volume}{12},
	\bibinfo{pages}{129--148}.
	%Type = Article
	\bibitem[{Oh and Patton(2017)}]{Oh.Patton2017}
	\bibinfo{author}{Oh, D.\, H.}, \bibinfo{author}{Patton, A.},
	\bibinfo{year}{2017}.
	\newblock \bibinfo{title}{Modeling dependence in high dimensions with factor
		copulas}.
	\newblock \bibinfo{journal}{Journal of Business and Economic Statistics}
	\bibinfo{volume}{35}, \bibinfo{pages}{139--154}.
	%Type = Article
	\bibitem[{Sklar(1959)}]{Sklar1959}
	\bibinfo{author}{Sklar, A.}, \bibinfo{year}{1959}.
	\newblock \bibinfo{title}{Fonctions de r\'epartition \`a $n$ dimensions et
		leurs marges}.
	\newblock \bibinfo{journal}{Institute of Statistics of the University of Paris}
	\bibinfo{volume}{8}, \bibinfo{pages}{229--231}.
	%Type = Book
	\bibitem[{Stroud and Secrest(1966)}]{Stroud.Secrest1966}
	\bibinfo{author}{Stroud, A.}, \bibinfo{author}{Secrest, D.},
	\bibinfo{year}{1966}.
	\newblock \bibinfo{title}{Gaussian {Q}uadrature {F}ormulas}.
	\newblock \bibinfo{publisher}{Prentice-Hall}, \bibinfo{address}{Englewood
		Cliffs, NJ}.
	%Type = Article
	\bibitem[{Yoshiba(2018)}]{Yoshiba2018}
	\bibinfo{author}{Yoshiba, T.}, \bibinfo{year}{2018}.
	\newblock \bibinfo{title}{Maximum likelihood estimation of skew-t copulas with
		its applications to stock returns}.
	\newblock \bibinfo{journal}{Journal of Statistical Computation and Simulation}
	\bibinfo{volume}{88(13)}, \bibinfo{pages}{2489--2506}.
	
\end{thebibliography}

\end{document}